\documentclass[useAMS,usenatbib]{mn2e}
\pdfoutput=1
\newcommand{\DIFdelFL}[1]{}
\newcommand{\DIFaddFL}[1]{}

\usepackage{graphicx, subfigure}
\usepackage{multicol}
\usepackage{bibentry}
\usepackage{lipsum}
\usepackage{array,booktabs,dcolumn,calc}
\newcolumntype{C}{>{\centering\arraybackslash}b{\widthof{positions}}}
\newcolumntype{d}{D{.}{.}{-2}}
\usepackage{pdfpages}

\usepackage{tikz}
\usepackage{float}
\usepackage{verbatim}
\usepackage{amssymb,amsmath}
\usepackage{setspace}
\usepackage{rotating,rotfloat}

\usepackage{wasysym}

\usepackage{xcolor}
\usepackage{caption}
\usepackage{capt-of}
\usepackage{multirow}
\usepackage{floatpag}
\graphicspath{ {./figures/} }
\usepackage{titlesec}

\titlespacing{\section}{0pt}{\parskip}{-\parskip}
\titlespacing{\subsection}{0pt}{\parskip}{-\parskip}
\titlespacing{\subsubsection}{0pt}{\parskip}{-\parskip}

\DeclareRobustCommand{\ion}[2]{%
\relax\ifmmode
\ifx\testbx\f@series
{\mathbf{#1\,\mathsc{#2}}}\else
{\mathrm{#1\,\mathsc{#2}}}\fi
\else\textup{#1\,{\mdseries\textsc{#2}}}%
\fi}

\usepackage[colorlinks, citecolor=black, linkcolor=blue]{hyperref}
\usepackage{cite}
\renewcommand{\thefootnote}{\roman{footnote}}

\usepackage[titletoc,toc,title]{appendix}
\usepackage{overpic}

\title[Dense circum-nuclear molecular gas in starburst galaxies]{Dense circum-nuclear molecular gas in starburst galaxies}
\author[Green et al. 2015]{C.-E. Green$^{1,2}$ $^{\star}$, M. R. Cunningham$^{1}$, J. A. Green$^{2,3}$, J. R. Dawson$^{2,5}$, P. A. Jones$^{1}$, \newauthor \'A. R. L\'opez-S\'anchez$^{4,5}$, L. Verdes-Montenegro$^{6}$, C. Henkel$^{7,8}$, W. A. Baan$^{9,10}$,  \newauthor S. Mart{\'i}n$^{11}$ \\ 
\\
$^{1}$School of Physics, University of New South Wales, Sydney, NSW, 2052, Australia \\
$^{2}$CSIRO Astronomy \& Space Science, Australia Telescope National Facility, PO Box 76, Epping, NSW 1710, Australia\\
$^{3}$SKA Organisation, Jodrell Bank Observatory, Lower Withington, Macclesfield SK11 9DL, UK \\
$^{4}$Australian Astronomical Observatory, PO Box 915, North Ryde, NSW 1670, Australia\\
$^{5}$Department of Physics and Astronomy and MQ Research Centre in Astronomy, Astrophysics and Astrophotonics, \\
Macquarie University, NSW 2109, Australia\\
$^{6}$Instituto de Astrof{\'i}sica de Andaluc{\'i}a (CSIC), Glorieta de la Astronom{\'i}a, s/n. E-18008, Granada, Spain\\
$^{7}$Max-Planck-Institut f{\"u}r Radioastronomie, Auf dem H{\"u}gel 69, 53121 Bonn, Germany\\
$^{8}$Astronomy Department, King Abdulaziz University, P.O. Box 80203, Jeddah 21589, Saudi Arabia\\
$^{9}$Shanghai Astronomical Observatory, Chinese Academy of Sciences, 80 Nandan Lu, Xuhui, Shanghai 200030, P.R. China\\
$^{10}$Netherlands Institute for Radio Astronomy, ASTRON, Oude Hoogeveensedijk 4, 7991 PD Dwingeloo, The Netherlands \\
$^{11}$Institut de Radio Astronomie Millim{\'e}trique, 300 rue de la Piscine, Dom. Univ., 38406, St. Martin d'H{\`e}res, France \\}
\begin{document}

\date{Placeholder for date 2015}

\pagerange{\pageref{firstpage}---\pageref{lastpage}} \pubyear{2015}

\maketitle

\label{firstpage}

\begin{abstract}
We present results from a study of the dense circum-nuclear molecular gas of starburst galaxies. The study aims to investigate the interplay between starbursts, active galactic nuclei and molecular gas. We characterise the dense gas traced by HCN, HCO$^{+}$ and HNC and examine its kinematics in the circum-nuclear regions of nine starburst galaxies observed with the Australia Telescope Compact Array. We detect HCN (1--0) and HCO$^{+}$ (1--0) in seven of the nine galaxies and HNC (1--0) in four.  Approximately 7\,arcsec resolution maps of the circum-nuclear molecular gas are presented. The velocity integrated intensity ratios, HCO$^{+}$ (1--0)/HCN (1--0) and HNC (1--0)/HCN (1--0), are calculated. Using these integrated intensity ratios and spatial intensity ratio maps we identify photon dominated regions (PDRs) in NGC~1097, NGC~1365 and NGC~1808. We find no galaxy which shows the PDR signature in only one part of the observed nuclear region. We also observe unusually strong HNC emission in NGC~5236, but it is not strong enough to be consistent with X-ray dominated region (XDR) chemistry. Rotation curves are derived for five of the galaxies and dynamical mass estimates of the inner regions of three of the galaxies are made. 
\end{abstract}


\section{Introduction}
\label{sec:introduction}
\let\thefootnote\relax\footnote{$^{\star}$E-mail: claire.elise.green@gmail.com} 
Understanding the relationship between star formation and active galactic nuclei (AGN) is central to the study of galaxy formation and evolution. Starbursts and AGN are known to coexist in many galaxies \citep[e.g.][]{b33, b84, b137}. Bulge properties of the host galaxy, such as spheroid luminosity \citep{b169}, bulge mass \citep{b168}, stellar velocity dispersion \mbox{\citep[e.g.][]{Tremaine2002, Woo2013}} and bulge concentration \citep{b167}, have been found to be tightly correlated to the mass of the central black hole, suggesting that the black hole and the host galaxy are closely related and may co-evolve. Starbursts are associated with spheroid formation \citep{barnes91} and since the spheroid properties are closely correlated with those of the AGN (or black hole) a starburst-AGN connection is to be expected. This connection may be evolutionary; AGN have been suggested as the final stage of nuclear starburst evolution \citep{weedman83}. The starburst may also regulate the amount of gas available for accretion onto the AGN \mbox{\citep{b224}} or the AGN may regulate circum-nuclear star formation or even quench it \citep[e.g.][]{Dubois2013, Olsen2013}. Alternatively these phenomena may be decoupled, occurring simultaneously because they are triggered and fuelled by common mechanisms: interactions, such as mergers, and a rich supply of circum-nuclear gas. \\

Although the precise nature of the starburst-AGN connection remains an open question, some progress in this field includes the correlation between star formation rate (SFR) and average black hole accretion rate \citep{chen2013}. With this project we seek to investigate the influence of star formation and AGN on circum-nuclear molecular gas, laying the foundation for the use of this gas in the study of the interplay between AGN and starbursts. \\

Molecular line intensity ratios can be used as diagnostic tools to examine the influence of the star formation or AGN on circum-nuclear gas (e.g. \citealt*{b214}; \citealt*{b185}; \citealt{b219}; \citealt{loenen2008}). These intensity ratios allow for the identification of X-ray dominated regions (XDRs) and photon dominated regions (PDRs), which facilitates determination of whether the star formation or AGN has the dominant effect on the gas \citep{b185}. PDRs are affected by star formation regions producing far-ultraviolet (FUV) radiation. XDRs are primarily influenced by X-rays from an AGN or black hole. The chemistry of these regions is dependent upon the type of radiation, thus they are differentiated by the relative intensity of their rotational molecular lines \citep{b185}. In our characterisation of the dense, circum-nuclear molecular gas we use velocity integrated flux density ratios (henceforth `integrated intensity ratios') to determine whether XDRs or PDRs are present. A point of difference of this project compared to others examining the same sources is the spatial information obtained by using telescope array data. This allows the creation of spatially resolved integrated intensity ratio maps that display the spatial variation of the ratios. These maps are an important diagnostic tool in the robust identification of XDRs and PDRs that can be used alongside traditional intensity ratio values. \\

The observations and data reduction are described in~\autoref{sec:data}; results are presented in~\autoref{sec:results_discussion} along with the discussion of the implications of these results for characterising the dense circum-nuclear molecular gas of these galaxies. Conclusions are presented in~\autoref{sec:conclusions}. \\

\section{Observations}
\label{sec:data}
\subsection{Sources}
\label{sec:sources}
\begin{figure*}
	\thisfloatpagestyle{empty}      
\includegraphics[trim=3cm 0cm 3cm 0cm,clip=true,scale=0.85]{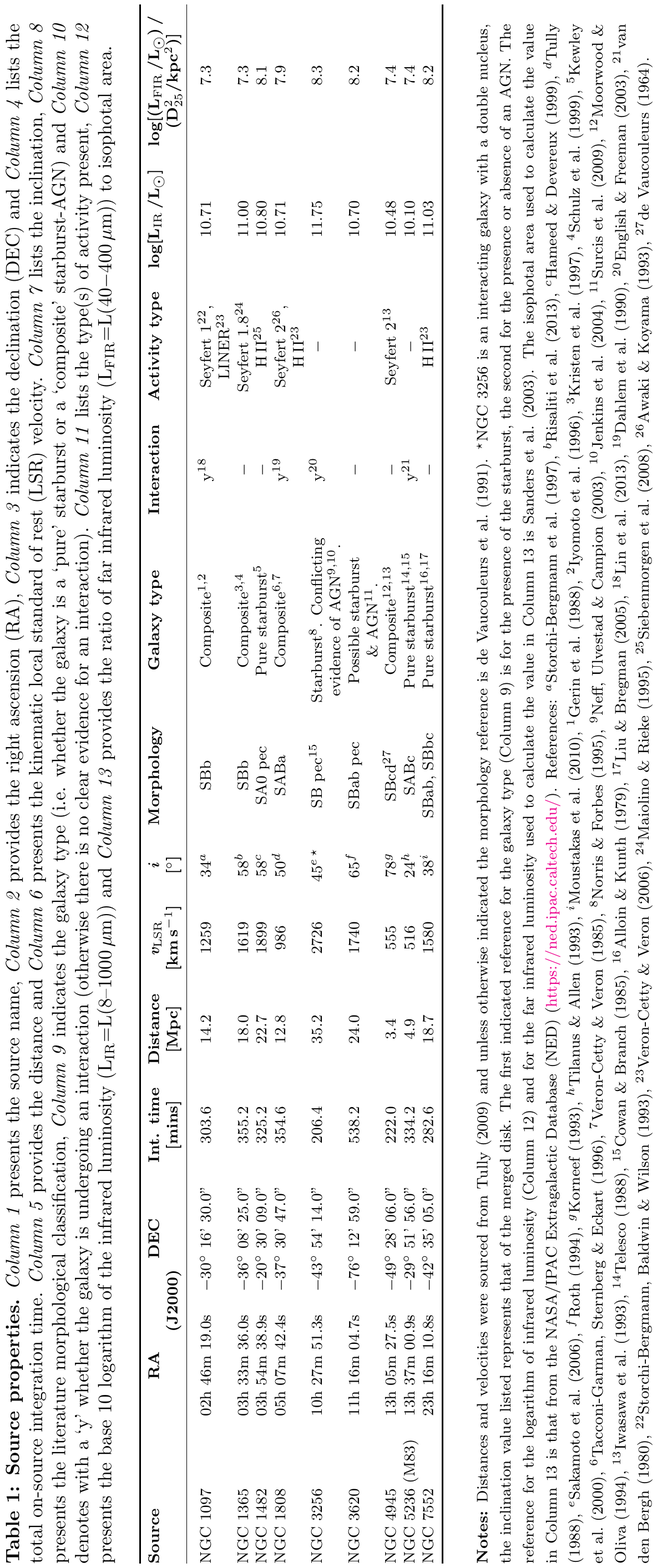}
 
\label{fig:table1}
\end{figure*}

Previously unprocessed data were retrieved from the Australia Telescope Online Archive (ATOA$^{1}$\let\thefootnote\relax\footnote{$^{1}$\href{http://atoa.atnf.csiro.au}{http://atoa.atnf.csiro.au}}). To achieve the aims outlined in~\autoref{sec:introduction} the archive was searched for resolved composite AGN/starburst and pure starburst sources with high resolution observations of HCN, HCO$^{+}$ and HNC (1--0). A sample of nine galaxies was selected according to these conditions from the dataset corresponding to project code C2116 (primary investigator D. Espada). The sources included in this dataset, listed in~\hyperref[fig:table1]{Table 1}, are southern infrared (IR) bright galaxies with strong CO (2--1) molecular lines. To probe the relationship between starbursts and AGN, we selected sources that are starbursts hosting either confirmed AGN, possible AGN or no AGN. All sources meet the starburst criteria of \citet{Mao2010}, where a starburst galaxy has a ratio of infrared luminosity (L$_{\rm{FIR}}$=L(40$-$400\,$\mu$m)) to isophotal area satisfying log[(L$_{\rm{FIR}}$/L$_{\astrosun}$)/(D$_{25}^{2}$/kpc$^{2}$)]\textgreater 7.25. The galaxies are all barred spirals with the exception of NGC~1482, which is a lenticular peculiar galaxy \citep{deVal91}. The morphology, luminosity and other properties of the sources are summarised in~\hyperref[fig:table1]{Table 1}. \\

The nine galaxies listed in~\hyperref[fig:table1]{Table 1} were observed with the hybrid H75 configuration of the Australia Telescope Compact Array (ATCA), located near Narrabri, NSW, Australia. These observations were undertaken in the period from July to October 2009. Observations were made periodically of each source and an appropriate phase calibrator, switching from the source to the phase calibrator, and then back to the source. Uranus was used as the flux calibrator throughout the observations. \\

The H75 array is the most compact array configuration available and was used to produce a synthesised beamsize of $\sim$7$\times$7\,arcsec. The Compact Array Broadband Backend (CABB$^{2}$\let\thefootnote\relax\footnote{$^{2}$\href{http://www.narrabri.atnf.csiro.au/observing/CABB.html}{http://www.narrabri.atnf.csiro.au/observing/CABB.html}}) provided 2$\times$2\,GHz bandwidth windows for the observations. Each window delivered dual polarisation data with 2048 channels. The central frequencies of the windows were 89\,GHz to observe the HCN (1--0) and HCO$^{+}$ (1--0) transitions occurring at rest frequencies of 88.63 and 89.19\,GHz respectively, and 91\,GHz to observe the HNC (1--0) transition occurring at a rest frequency of 90.66\,GHz. The data have a frequency resolution of 1\,MHz, corresponding to velocity channel resolutions of 3.4 and 3.2\,km\,s$^{-1}$ for the lower and upper windows respectively, and spectral bandwidths of 6900 and 6750\,km\,s$^{-1}$. \\

\setlength\fboxsep{0pt}
\setlength\fboxrule{0.5pt}
\setcounter{table}{1}

\subsection{Data reduction}
The unprocessed data files were obtained from the ATOA and data reduction was performed using \textsc{miriad}$^{3}$\let\thefootnote\relax\footnote{$^{3}$\href{http://www.atnf.csiro.au/computing/software/miriad/}{http://www.atnf.csiro.au/computing/software/miriad/}} (version 1.5; \citealt{b16}), following the standard procedure, to produce spectral line data cubes from data collected across multiple epochs in 2009. \\

Molecular line datasets for each frequency were calibrated separately, then combined (using \textsc{miriad} \textsc{invert}) to produce one data cube per molecular line per source. A primary beam correction was performed using the \textsc{miriad} \textsc{linmos} task. The H{\"o}gbom \textsc{clean} algorithm \citep{hog74} was applied to deconvolve the data within \textsc{miriad}. The cleaned HCO$^{+}$ and HNC data were re-gridded to the HCN spatial grid. \\

Area integrated spectra were produced for each molecular line detected in each source. To use these spectra later in the calculation of integrated intensity ratios, we required the best possible signal-to-noise. We chose to produce these spectra over the same area for all lines, which was the area that included all significant emission. To achieve this the spectra were produced over the area defined by the $\geq$3$\sigma$ contour of the strongest line. This line provided the largest area of the three, and this $\geq$3$\sigma$ region is referred to as the `largest emission region'. Integrated spectra (as opposed to average spectra) were produced so that any pixels with no emission in the other two lines did not contribute to the spectra. These spectra can then be directly compared in the calculation of integrated intensity ratios. The largest emission region was that of HCN for NGC~1097, NGC~1365 and NGC~1808, and HCO$^{+}$ for NGC~3256, NGC~4945, NGC~5236 and NGC~7552.  \\

Spectra were produced by first constructing a 2D mask from the moment zero (velocity integrated specific intensity) map of the molecular line that had the largest $\geq$3$\sigma$ emission region. This mask was then applied to each velocity slice of the 3D data cube of each molecular line detected and the unmasked pixel values were summed and converted from units of specific intensity (Jy\,beam$^{-1}$) to units of flux density (Jy) by multiplying by the pixel area in beam units to produce the integrated flux density value plotted against the velocity$^{4}$\let\thefootnote\relax\footnote{$^{4}$All velocities presented are kinematic local standard of rest (LSR) velocities.} values in the final integrated spectra presented in~\autoref{fig:av_spectra}. The average rms of the data cubes with detections was 8.1\,mJy, for cubes with non-detections the average rms was 13.4\,mJy. Poor weather, which has a particularly adverse effect on 3$-$mm observations, contributed to this difference in the rms of the detections and non-detections. \\

Gaussian fitting of the integrated spectra using $\chi^{2}$ minimisation was performed with the \textsc{python} \textsc{mpfit}$^{5}$\let\thefootnote\relax\footnote{$^{5}$Sergey Koposov's \textsc{python} translation of \textsc{mpfit}: \href{https://code.google.com/p/astrolibpy/downloads/list}{https://code.google.com/p/astrolibpy/downloads/list}} routine to extract the peak flux density, central velocity, linewidth (FWHM) and velocity integrated flux density (Jy\,km\,s$^{-1}$, henceforth `integrated flux density'). The spectra were Hanning smoothed in \textsc{python} with window sizes of 3, 5, 7 or 9 velocity channels as appropriate before Gaussian fitting was performed.\\

\section{Results and discussion}
\label{sec:results_discussion}
	\subsection{HCN, HCO\texorpdfstring{\textsuperscript{+}} \enspace \enspace and HNC detections}
	\label{sec:HCN, HCO and HNC Detections}
\begin{figure*}
	\caption[HCN, HCO$^{+}$ and HNC integrated spectra]{\textbf{HCN, HCO$^{+}$ and HNC integrated spectra.} Source name and molecular species (J=1$\rightarrow$0 transition) are given on each spectrum. The dashed line represents the Gaussian fit, whose parameters are listed in~\autoref{tab:fit_params_new}.}

\includegraphics[trim=0cm 0cm 0cm 0cm,clip=true,scale=0.79,angle=90]{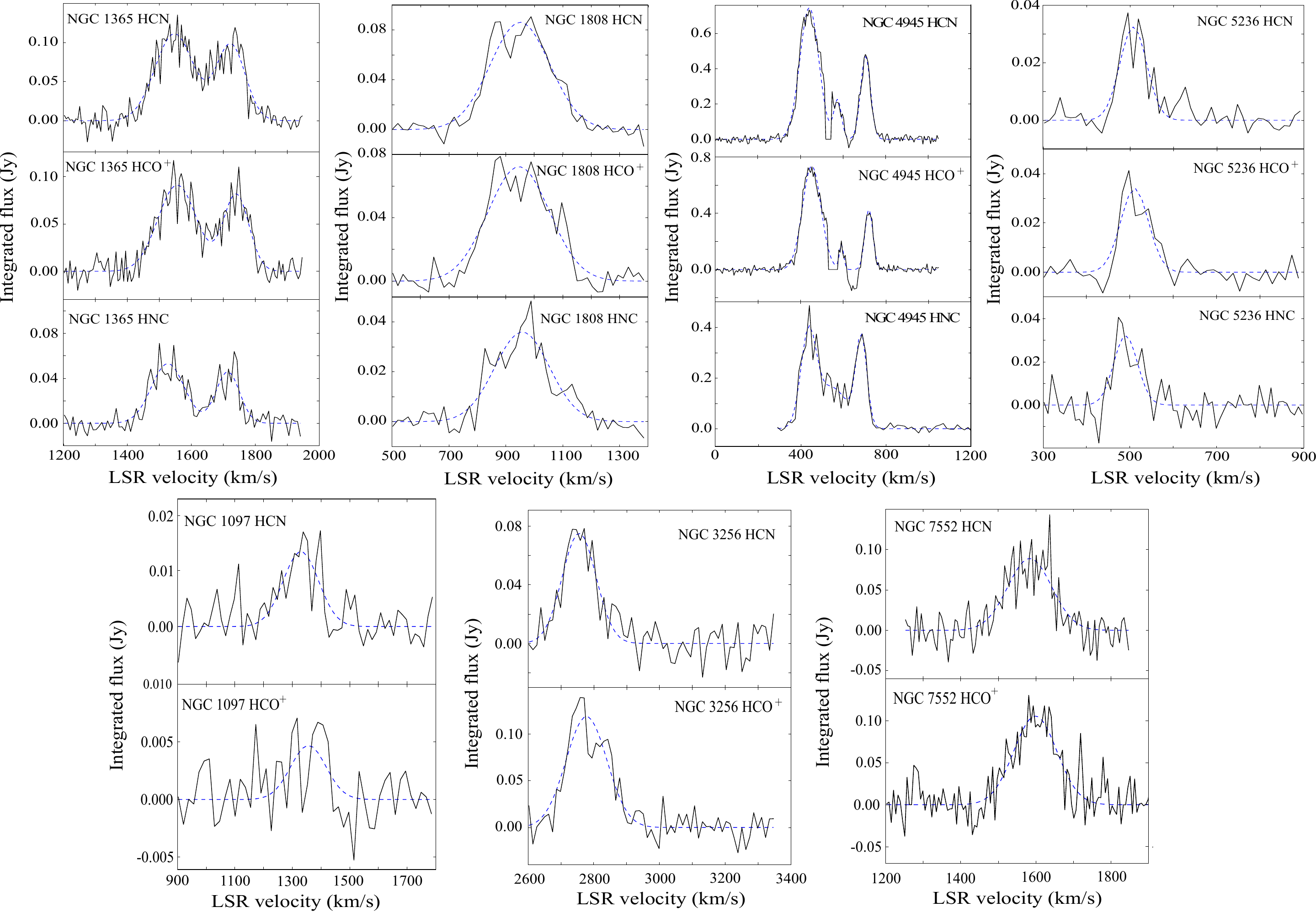}
 \label{fig:av_spectra}
\end{figure*}

\begin{figure*}
\begin{center}
\centering
\includegraphics[trim=2.5cm 7.8cm 2.5cm 7.5cm,clip=true,width=0.9\linewidth]{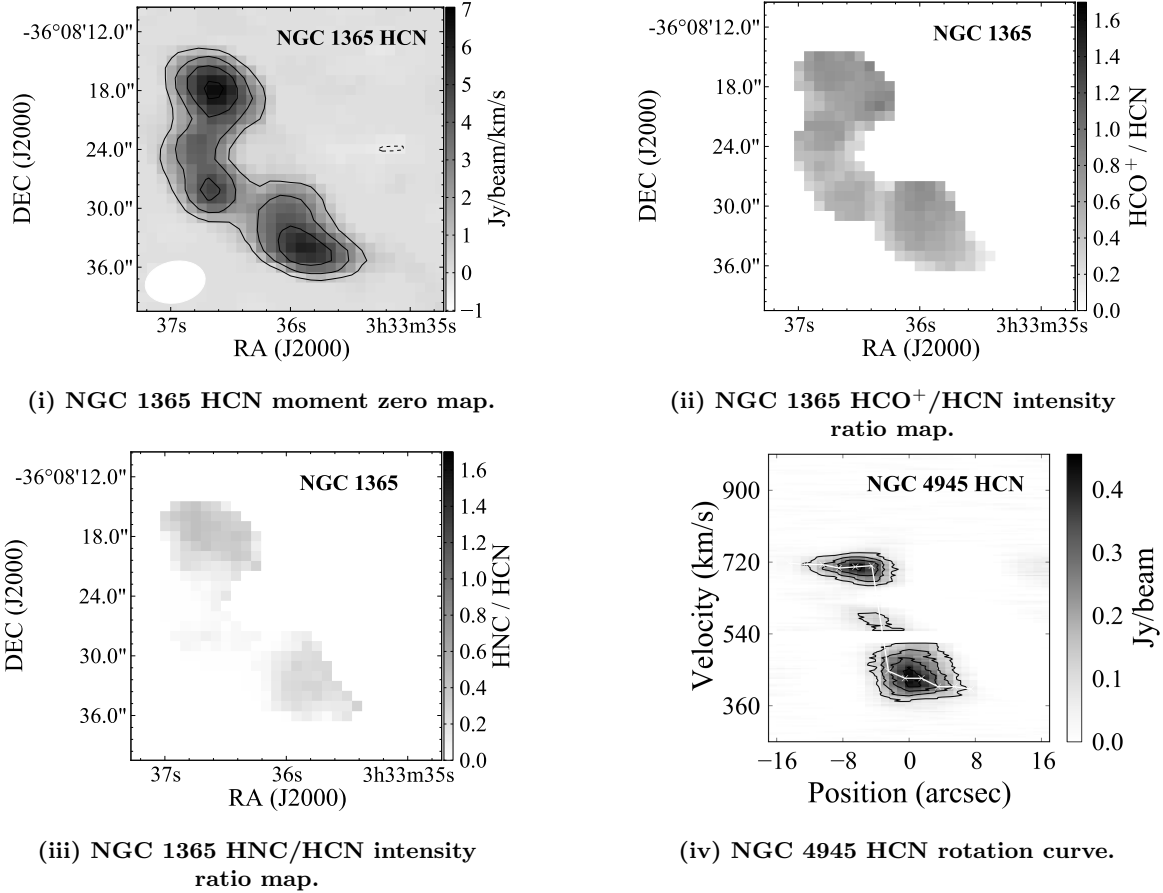}
\caption{\textbf{Example plots.} Example moment zero (velocity integrated specific intensity) map, intensity ratio maps, position-velocity diagram and rotation curve. The complete set of these plots for each relevant source can be found in Online Appendices A, B and C respectively.}
\label{fig:example_plots}
\end{center}
\end{figure*}

	\noindent Detections of HCN (1--0) and HCO$^{+}$ (1--0) have been made in seven of the targeted galaxies and HNC (1--0) has been detected in four. The integrated spectra for the molecular lines of each galaxy are presented in~\autoref{fig:av_spectra}. The parameters of the Gaussian fit: peak flux density, central velocity, linewidth and integrated flux density, are presented in~\autoref{tab:fit_params_new} for HCN, HCO$^{+}$ and HNC respectively. Moment zero maps of the dense gas tracers are presented in Online Appendix A. The moment zero maps of the NGC~1365 molecular lines are presented in~\autoref{fig:example_plots} as an example. Upper limits for the non-detections are presented in~\autoref{tab:fit_params_new}. The upper limit on the peak flux density was taken to be three times the rms of the non-detection data cube, while the upper limit of the integrated flux density was taken to be three times the rms of the moment zero map of the non-detection. For sources with non-detections of all three lines (NGC~1482 and NGC~3620) these measurements were made over the CO (1$-$0) velocity range of the literature \citep{Elfhag1996}. For sources with a non-detection of HNC (NGC~1097, NGC~3256 and NGC~7552) these measurements were made over the approximate velocity range encompassing the region where HCN and HCO$^{+}$ were detected.\\

In NGC~4945 strong detections of HCN, HCO$^{+}$ and HNC have been made. The molecular lines appear to have two major components at $\sim$440\,km\,s$^{-1}$ and $\sim$700\,km\,s$^{-1}$ and a weaker component at $\sim$570\,km\,s$^{-1}$. The HCN and HCO$^{+}$ spectra dip below zero at $\sim$625\,km\,s$^{-1}$. This is likely due to absorption against the 3$-$mm continuum. \citet{b201} also observed this absorption feature at the same approximate velocity. The `flat' sections of the HCN and HCO$^{+}$ spectra at $\sim$530\,km\,s$^{-1}$ are due to clipping in the data reduction to remove large noise spikes that were likely from interference. \\
	
To approximate the amount of emission on more extended spatial scales filtered out in the ATCA interferometric observations (the `missing flux') we compare$^{6}$\let\thefootnote\relax\footnote{$^{6}$To calculate the recovered flux we divided the ATCA integrated flux by the \citet{b219} SEST integrated flux and convert this to a percentage.} with available single dish Swedish-ESO Submillimetre Telescope (SEST) observations of HCN, HCO$^{+}$ and HNC presented in \citet{b219}. Excluding NGC~5236, where percentages are lower, we recover between 19\% and 50\% of the flux, with an average of 25\%. The amount of recovered flux is listed for each source in~\autoref{tab:fit_params_new}. \\

We explore the potential effect of the missing flux on our analysis in Online Appendix B. The amount of missing flux for each molecular line in each source is estimated, and  intensity ratios adjusted for this missing flux are also calculated (please see Online Appendix B for full details). We find the adjusted ratios to be consistent with the ratios calculated in~\autoref{sec:calc_of_ratios}, generally differing from these ratios by less than 10\%. The missing flux therefore does not present a significant problem, not even in the case of NGC~5236. We find that in practice the scientific conclusions of this work are robust to the amount of missing flux over the small areas we are considering.\\

Interferometric data have advantages for the calculation of integrated intensity ratios between different molecular lines, both due to the higher baseline stability and higher angular resolution. This allows direct comparison of the molecular line ratios in spatially resolved regions within the central region of galaxies, which are averaged at the single dish resolution. Our ratios are not global values for the galaxy, but are representative of the conditions of localised density enhancements on a scale of $\sim$7\,arcsec within a more extended structure. With the interferometric data we can then produce ratio maps showing the spatial variation of the ratios on these scales. \\

\setlength\fboxsep{0pt}
\setlength\fboxrule{0.5pt}

\begin{table*} 
\begin{center}

\caption[Detections]{\textbf{HCN, HCO$^{+}$ and HNC (1--0) detections and non-detections.} Listed are the parameters of the Gaussian fits of the seven HCN (1--0) and HCO$^{+}$ (1--0) detections and the four HNC (1--0) detections. In the case of a non-detection the 3$\sigma$ upper limits for the flux density and integrated flux density are listed. These values represent three times the rms of the data cube (Col. 2) and three times the rms of the moment zero map (Col. 5) respectively. The width of an individual velocity channel is also listed for the non-detections (Col. 4). \emph{Column 1} is the source name, \emph{Column 2} lists the peak flux density after the primary beam correction, \emph{Column 3} provides the kinematic LSR velocity of the peak of the integrated spectra, \emph{Column 4} gives the linewidth (FWHM), \emph{Column 5} presents the velocity integrated flux density and \emph{Column 6} presents an estimate of the recovered flux of the interferometric data compared the the single-dish data of \citet{b219}.} 
\label{tab:fit_params_new}

\begin{tabular}{@{\qquad \quad} l r r r r c @{}}
	&		&		&		&	&	\\
\toprule

\textbf{Source}	&	\multicolumn{1}{c}{\textbf{S}}	&	\multicolumn{1}{c}{\textbf{\emph{v}$_{\rm{LSR}}$}}	&	\multicolumn{1}{c}{\textbf{FWHM}}	&	\multicolumn{1}{c}{\textbf{$\int$ S dv}} &		\multicolumn{1}{c}{\textbf{Recovered}}	\\
	&	\multicolumn{1}{c}{\textbf{[mJy]}}	&	\multicolumn{1}{c}{\textbf{[km\,s$^{-1}$]}}	 &	\multicolumn{1}{c}{\textbf{[km\,s$^{-1}$]}}	&	\multicolumn{1}{c}{\textbf{[Jy\,km\,s$^{-1}$]}}	& \multicolumn{1}{c}{\textbf{flux [\%]}}	\\
	\midrule
\multicolumn{5}{@{}l}{\textbf{HCN}}			\\
NGC~1097	&	13.0	\,$\pm$\,	1.5	&	1329.9	\,$\pm$\,	8.5	&	146.1	\,$\pm$\,	20.0	&	2.0	\,$\pm$\,	0.3	& $-$ \\
NGC~1365	&	111.1	\,$\pm$\,	2.9	&	1547.2	\,$\pm$\,	2.2	&	143.0	\,$\pm$\,	5.7	&	16.9	\,$\pm$\,	0.8	& 21 \\
	&	97.0	\,$\pm$\,	3.3	&	1723.5	\,$\pm$\,	2.2	&	107.2	\,$\pm$\,	5.3	&	11.1	\,$\pm$\,	0.6	\\
NGC~1482	&	\textless34.0		& $-$ &	3.4 &	\textless1.8 &	$-$ \\
NGC~1808	&	86.2	\,$\pm$\,	1.8	&	950.6	\,$\pm$\,	2.6	&	247.3	\,$\pm$\,	6.0	&	22.7	\,$\pm$\,	0.7	 & 21 \\
NGC~3256	&	75.1	\,$\pm$\,	5.0	&	2755.5	\,$\pm$\,	3.9	&	121.2	\,$\pm$\,	9.3	&	9.7	\,$\pm$\,	0.9	& 21 \\
NGC~3620	&	\textless36.1		& $-$ &	3.0 &		\textless5.3	& \textless13\\
NGC~4945	&	745.6	\,$\pm$\,	4.8	&	441.9	\,$\pm$\,	0.3	&	105.0	\,$\pm$\,	0.8	&	83.4	\,$\pm$\,	0.8	& 29\\
	&	218.1	\,$\pm$\,	7.4	&	574.9	\,$\pm$\,	0.7	&	42.3	\,$\pm$\,	1.7	&	9.8	\,$\pm$\,	0.5	\\
	&	469.1	\,$\pm$\,	6.5	&	705.7	\,$\pm$\,	0.4	&	54.9	\,$\pm$\,	0.9	&	27.4	\,$\pm$\,	0.5	\\
NGC~5236	&	32.4	\,$\pm$\,	2.0	&	507.3	\,$\pm$\,	2.2	&	73.0	\,$\pm$\,	5.2	&	2.5	\,$\pm$\,	0.2	& 4 \\
NGC~7552	&	88.8	\,$\pm$\,	5.1	&	1583.8	\,$\pm$\,	3.9	&	138.0	\,$\pm$\,	9.2	&	13.0	\,$\pm$\,	0.1	& 38 \\

	&				&				&				&				\\
\multicolumn{5}{@{}l}{\textbf{HCO$^{+}$}}			\\
NGC~1097	&	4.7	\,$\pm$\,	1.0	&	1356.1	\,$\pm$\,	15.1	&	145.1	\,$\pm$\,	35.5	&	0.7	\,$\pm$\,	0.2	& $-$ \\
NGC~1365	&	91.0	\,$\pm$\,	2.9	&	1556.4	\,$\pm$\,	2.3	&	142.9	\,$\pm$\,	6.0	&	13.8	\,$\pm$\,	0.7	& 25 \\
	&	81.0	\,$\pm$\,	3.5	&	1740.8	\,$\pm$\,	2.2	&	93.8	\,$\pm$\,	5.2	&	8.1	\,$\pm$\,	0.5	\\
NGC~1482	&	\textless34.0	& $-$ &	3.4 &	\textless1.8	& $-$ \\
NGC~1808	&	72.2	\,$\pm$\,	2.1	&	945.9	\,$\pm$\,	3.6	&	254.2	\,$\pm$\,	8.5	&	19.5	\,$\pm$\,	0.9	& 33 \\
NGC~3256	&	118.8	\,$\pm$\,	5.3	&	2778.2	\,$\pm$\,	3.2	&	146.3	\,$\pm$\,	7.6	&	18.5	\,$\pm$\,	1.2	& 31 \\
NGC~3620	&	\textless41.3	& $-$ &	3.0 &		\textless5.0	& \textless22 \\
NGC~4945	&	735.9	\,$\pm$\,	5.3	&	452.0	\,$\pm$\,	0.4	&	102.2	\,$\pm$\,	0.8	&	80.1	\,$\pm$\,	0.8	& 23 \\
	&	172.0	\,$\pm$\,	10.8	&	591.8	\,$\pm$\,	0.7	&	24.0	\,$\pm$\,	1.7	&	4.4	\,$\pm$\,	0.4	\\
	&	426.6	\,$\pm$\,	8.0	&	720.3	\,$\pm$\,	0.4	&	44.4	\,$\pm$\,	1.0	&	20.1	\,$\pm$\,	0.5	\\
NGC~5236	&	34.1	\,$\pm$\,	2.2	&	511.1	\,$\pm$\,	2.4	&	74.9	\,$\pm$\,	5.6	&	2.7	\,$\pm$\,	0.3	& 9 \\
NGC~7552	&	105.6	\,$\pm$\,	6.6	&	1598.6	\,$\pm$\,	4.1	&	134.7	\,$\pm$\,	9.7	&	15.2	\,$\pm$\,	1.4	& 50 \\

&		&		&		&		\\

\multicolumn{5}{@{}l}{\textbf{HNC}}			\\
NGC~1097	&	\textless43.3	& $-$ &	3.0 &		\textless1.0	& $-$ \\
NGC~1365	&	53.1	\,$\pm$\,	2.6	&	1526.4	\,$\pm$\,	2.9	&	121.6	\,$\pm$\,	7.2	&	6.9	\,$\pm$\,	0.5	& 23 \\
	&	45.8	\,$\pm$\,	3.0	&	1713.9	\,$\pm$\,	5.4	&	86.7	\,$\pm$\,	6.9	&	4.2	\,$\pm$\,	0.4	\\
NGC~1482	&	\textless24.8		& $-$ &	3.3 &	\textless1.7	& $-$ \\
NGC~1808	&	36.0	\,$\pm$\,	1.2	&	957.8	\,$\pm$\,	3.7	&	226.9	\,$\pm$\,	8.8	&	8.7	\,$\pm$\,	0.4	& 19 \\
NGC~3256	&	\textless78.9	& $-$ &	3.0 &	\textless5.5 &	\textless13 \\
NGC~3620	&	\textless28.0	& $-$ &	3.3 &	\textless2.3	&	\textless15 \\
NGC~4945	&	348.0	\,$\pm$\,	138.1	&	435.0	\,$\pm$\,	4.8	&	83.9	\,$\pm$\,	19.8	&	31.1	\,$\pm$\,	13.5	& 38 \\
	&	163.3	\,$\pm$\,	24.1	&	543.6	\,$\pm$\,	48.0	&	177.0	\,$\pm$\,	135.6	&	30.8	\,$\pm$\,	22.6	\\
	&	355.6	\,$\pm$\,	53.2	&	683.2	\,$\pm$\,	2.8	&	59.4	\,$\pm$\,	9.2	&	22.5	\,$\pm$\,	4.5	\\
NGC~5236	&	32.0	\,$\pm$\,	3.2	&	490.5	\,$\pm$\,	3.2	&	64.7	\,$\pm$\,	7.5	&	2.2	\,$\pm$\,	0.3	& 11 \\
NGC~7552	&	\textless42.3	& $-$ &	3.0 &	\textless2.9 & 	\textless18 \\

\bottomrule

\end{tabular}
\end{center}
\end{table*}

We find the HCO$^{+}$ largely follows the HCN distribution which is to be expected based on what is seen in our own Galaxy \citep{b238, Jones2012}. In NGC~5236 the HNC line approaches the intensity of the HCN line, which is unusual. Although these molecules are all dense gas tracers, HNC generally traces cool gas as it is preferentially formed over HCN at temperatures $\lesssim$15\,K, while HCN and HCO$^{+}$ trace warmer gas. \citet*{b189} and \citet{Aalto07} investigated anomalously bright HNC comparable to the HCN emission. They suggest four possible explanations: large masses of hidden cold gas, ion-neutral dominated chemistry, XDR chemistry and HNC enhancement through mid-IR pumping. As will be discussed in detail in the following section, the integrated intensity ratios of NGC~5236 are not consistent with XDR chemistry. Our data do not exclude any of the three other possible explanations and chemical modelling could be used to investigate the anomalous HNC further. The modelling, however, is beyond the scope of this work. \\

\subsection{Integrated intensity ratios}		
\label{sec:intensity_ratios}

Recent numerical models of the heating effects of X-ray and UV radiation to produce XDR and PDR regions by \citet{meij05} and \citet{b185}, and the extension of these models to include mechanical heating by \citet{loenen2008}, demonstrate that the molecular line intensity ratios of \mbox{HCO$^{+}$ (1--0)/HCN (1--0)} and \mbox{HNC (1--0)/HCN (1--0)} discriminate between XDRs and PDRs. These molecular lines then allow us to determine whether the starburst or the AGN predominantly affects the chemistry of the molecular gas. These ratios will henceforth be referred to as HCO$^{+}$/HCN and HNC/HCN respectively. The predictions of these models will be described here and we compare our observational results with the XDR/PDR modelling in~\autoref{sec:ident_xdr_pdr_with_ratio}.\\

In general these models are anchored to the hydrogen number density. Three main models were presented by \citet{meij05} and \citet{b185} covering high (10$^{4}$--10$^{6.5}$\,cm$^{-3}$), moderate (10$^{3}$--10$^{4}$\,cm$^{-3}$) and low (10$^{2}$--10$^{3}$\,cm$^{-3}$) number densities. HCN, HCO$^{+}$ and HNC (1--0) are widely considered high density gas tracers and their critical densities$^{7}$\let\thefootnote\relax\footnote{$^{7}$Critical densities for HCN, HCO$^{+}$ and HNC (1$-$0) were calculated at a kinetic temperature of 100\,K using the equation: n(H$_{2}$)=A$_{ij}$/C$_{ij}$, where i=1 and j=0, with coefficients from the LAMBDA$^{6}$ database \citep{schoier2005} (\href{http://home.strw.leidenuniv.nl/~moldata/}{http://home.strw.leidenuniv.nl/$\sim$moldata/}).} (of $\sim$3$\times$10$^{6}$, $\sim$4$\times$10$^{5}$ and $\sim$4$\times$10$^{6}$\,cm$^{-3}$ respectively), when used as a first order approximation to the number density, indicate the high density XDR/PDR model is applicable. These critical densities hold for optically thin lines. In the case of optically thick conditions these critical densities will decrease with the inverse of the opacity. Deviations from optically thin conditions will shift line intensity ratios closer to unity, but ratios will remain to be either larger or smaller than one. It is likely these lines are sub-thermally excited, and that the emission arises from regions below the critical density. \citet{meij05} and \citet{b185} state that in the case of the HCN, HCO$^{+}$ and HNC molecular lines, only the highest density model (\textgreater 10$^{4}$\,cm$^{-3}$) is applicable due to the very poor observational prospects of these molecules in their lower density models. We therefore compare our ratios with the high density model results, following \citet{b219}. \\ 

Within the high density regime of \citet{meij05} and \citet{b185}:
\begin{enumerate}
\item XDRs are characterised by HCO$^{+}$/HCN and HNC/HCN ratios \textgreater 1 for high hydrogen number densities ($\geq$10$^{5}$\,cm$^{-3}$) and high hydrogen column densities (\textgreater10$^{23}$\,cm$^{-2}$). \\

\item PDRs are characterised by HCO$^{+}$/HCN and HNC/HCN ratios $\lesssim$1 in these high (column) density regions.
\end{enumerate}

Cosmic rays can also influence the chemistry of molecular gas, producing cosmic ray dominated regions (CRDRs). These regions can be detected in high J transition (J\,\textgreater\,10) CO lines. However, they are indistinguishable from XDRs in HCN, HCO$^{+}$ and HNC \citep{meij2011}. Similar to X-rays, the cosmic rays can enhance HCO$^{+}$ by increasing the ionisation of the gas, producing high HCO$^{+}$/HCN ratios \citep{papa2010}. This ambiguity is not an issue in this work. As will be discussed in~\autoref{sec:ident_xdr_pdr_with_ratio} no definitive XDR detections were made. \\

HCN, HCO$^{+}$ and HNC ratios have also been linked to the heating budget of different stages of starburst evolution \mbox{\citep{loenen2008, baan2010}}. Mechanical heating from Young Stellar Objects (YSOs) or supernovae influence the relative abundances of these species. Heating of the gas \mbox{(\textgreater 100\,K)} by mechanical processes can cause the conversion of HNC into HCN, decreasing the HNC/HCN ratio. \mbox{\citet{loenen2008}} proposes a division of PDRs into two groups: high density \mbox{($\geq$10$^{5}$\,cm$^{-3}$)} UV dominated PDRs and lower density \mbox{($\sim$10$^{4-5}$\,cm$^{-3}$)} mechanical feedback dominated PDRs. The higher density UV dominated PDRs are proposed to have a HNC/HCN ratio approaching a value of one and weak HCO$^{+}$ representing an early stage of star formation while the heating of mechanical feedback PDRs is dominated by shocks, lowering the HNC/HCN ratio, representing a later stage of star formation where some stars have generated supernova.\\

\subsubsection{Calculation of integrated intensity ratios}
\label{sec:calc_of_ratios}
\begin{figure}
\caption[Line ratios]{\textbf{Integrated intensity ratios.} Logarithmic HNC (1--0)/HCN (1--0) velocity integrated intensity ratio and the logarithmic HCO$^{+}$ (1--0)/HCN (1--0) ratio. 1$\sigma$ error bars are plotted in black. Each source has been assigned a separate symbol. Where two ratios have been calculated for one source for the lower and higher velocity Gaussian components, these have both been plotted and indicated in the legend as `R1' and `R2' respectively. Arrows indicate the logarithmic HNC/HCN ratio upper limit for sources with HNC non-detections. The dashed grey lines separate the regions where XDR and PDR chemistry dominate according to \mbox{\citet{meij05} and \citet{b185}}.}
\includegraphics[trim=1.2cm 0cm 3.2cm 1cm,clip=true,scale=0.5]{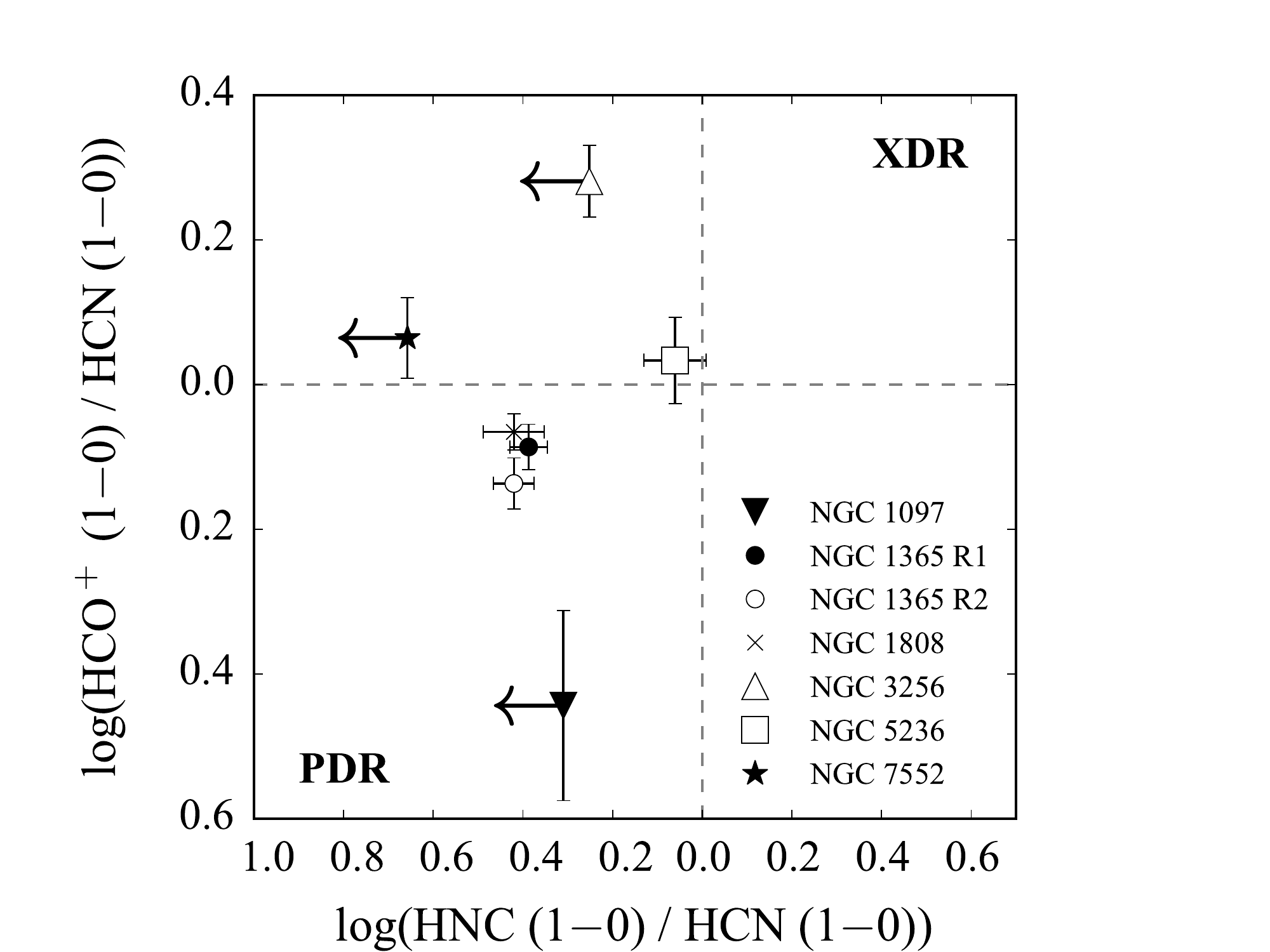}
\label{fig:line_ratio_plot}
\end{figure}

Integrated intensity ratios were calculated by division of the velocity integrated flux density of the individual Gaussian spectral components (see ~\autoref{fig:av_spectra}). Individual Gaussian components were treated separately as they are likely coming from spatially separate regions in the galaxy. The resulting integrated intensity ratio results are summarised in~\autoref{tab:line_ratio_results} and plotted in~\autoref{fig:line_ratio_plot}. As NGC~4945 was significantly affected by absorption features (see~\autoref{sec:results_discussion}) it is excluded from this analysis.\\

The integrated intensity ratios of the high density gas tracers HCN, HCO$^{+}$ and HNC are assumed to be unaffected by potential differences in their beam filling factors. These molecular lines have been observed with similar synthesised beams so under the assumption of similar spatial distribution their beam filling factors will also be much the same.\\

Intensity ratio maps were created by division of the relevant moment zero maps and are presented in Online Appendix C. These maps have been masked at the 3$\sigma$ level of the HCN moment zero map for each individual source. Example intensity ratio maps are presented for NGC 1365 in\mbox{~\autoref{fig:example_plots}}. The ratio maps display the spatial distribution and variation of the ratios and are important diagnostic tools. We urge caution in the interpretation of intensity ratio maps however, as they give no indication of the magnitude of the uncertainties. The numerical ratio with its standard deviation, while providing no indication of the size of spatially distinct chemical regions or of spatial variations of the ratio, does, however, provide a simple estimate of the uncertainty. We therefore encourage the employment of numerical intensity ratios and spatial ratio maps in parallel to ameliorate the process of robustly identifying XDRs and PDRs.\\

\begin{table} 
\begin{center}
\caption[Integrated intensity ratio results]{\textbf{Integrated intensity ratio results.} The HCO$^{+}$/HCN and HNC/HCN integrated intensity ratio results for the J=1$\rightarrow$0 transitions.  \emph{Column 1} is the source name. \emph{Column 2} provides the velocity integrated intensity ratio of Gaussian components corresponding in velocity. Where two ratios are listed, the first refers to the ratio of the first (lowest velocity) Gaussian components of the relevant spectra, while the second listing refers to the ratio of the second (higher velocity) Gaussian components.}
\label{tab:line_ratio_results}

\begin{tabular}{@{\qquad \quad} l l@{}}
\toprule
\textbf{Source}		&	\textbf{Ratio value}\\
	\midrule
	\noalign{\vskip 1mm}
\multicolumn{2}{@{}l}{\textbf{HCO$^{+}$/HCN}}			\\
NGC~1097	&	0.36	\,$\pm$\,	0.11	\\
NGC~1365	&	0.82	\,$\pm$\,	0.06	\\
	&	0.73	\,$\pm$\,	0.06	\\
NGC~1808	&	0.86	\,$\pm$\,	0.05	\\
NGC~3256	&	1.91	\,$\pm$\,	0.22	\\
NGC~5236	&	1.08	\,$\pm$\,	0.15	\\
NGC~7552	&	1.16	\,$\pm$\,	0.15	\\

	&				\\
\multicolumn{2}{@{}l}{\textbf{HNC/HCN}}			\\
NGC~1097	&		\textless0.49	\\
NGC~1365	&	0.41	\,$\pm$\,	0.04	\\
	&	0.38	\,$\pm$\,	0.04	\\
NGC~1808	&	0.38	\,$\pm$\,	0.06	\\
NGC~3256	&	\textless0.56	\\
NGC~5236	&	0.87	\,$\pm$\,	0.14	\\
NGC~7552	&	\textless0.22	\\

\bottomrule

\end{tabular}
\end{center}

\end{table}

\subsubsection{Identification of XDRs/PDRs with intensity ratios}
\label{sec:ident_xdr_pdr_with_ratio}
To make robust XDR/PDR classifications we require the calculated integrated intensity ratio to lie entirely above or below unity at the 1$\sigma$ uncertainty level and for the source to have spatially consistent ratio regions as demonstrated by the intensity ratio maps.\\

\citet{costa11} argue that the HCO$^{+}$/HCN ratio is not a reliable tracer of XDR/PDR chemistry when used alone, and \citet{b219} propose the HNC/HCN ratio is a much more dependable tracer of these regions compared to HCO$^{+}$/HCN. We therefore require both ratios to be consistent with XDR/PDR chemistry to make a determination. Three of the nine sources fully meet these conditions. Those galaxies providing a classification are described in more detail below.\\

\noindent\textbf{Details of XDR/PDR classifications:}\\
\noindent\textbf{NGC~1097:} The HCO$^{+}$/HCN ratio map has a value less than one. We propose NGC~1097 as a  PDR host. This is consistent with the suggestion by \citet{mason2007} that the dust component in the nuclear region of this galaxy is heated primarily by star clusters in the starburst ring.  \\

\citet{Martin2015} achieve higher resolution for this source with Atacama Large Millimeter/submillimeter Array (ALMA) observations and find variation in the HCO$^{+}$/HCN (1--0) ratio between the starburst ring and central region, with the HCO$^{+}$/HCN ratio decreasing with distance from the central AGN and distance to the starburst ring as would be expected in the scenario described by \citet{meij05} and \citet{b185}, summarised in~\autoref{sec:intensity_ratios}. They suggest that the AGN only affects the ratios at the very centre of the galaxy. Our overall ratios of less than one may be due to a PDR contribution from the starburst ring dominating over an XDR contribution coming from the central AGN. \citet{b185} suggest that the effect of an XDR is more difficult to detect if there is even a 10\% contribution from a PDR in the vicinity as this can suppress emission lines that would otherwise be enhanced. This means firm XDR detections may only be possible at very high resolution. \\

\noindent\textbf{NGC~1365:} The HCO$^{+}$/HCN and HNC/HCN ratio maps have values less than one. We conclude that NGC~1365 hosts a PDR, in agreement with the results of \mbox{\citet{b189}}, \mbox{\citet{b219}} and \mbox{\citet{baan2010}}. Our result represents the first use of high resolution intensity ratio maps to present a classification for this source. The lower HNC/HCN ratio value compared to the HCO$^{+}$/HCN ratio value indicates that NGC~1365 may host a PDR of the mechanical feedback dominated class (see Figure 1 b) of \citet{loenen2008}). \\

\noindent\textbf{NGC~1808:} The high resolution HCO$^{+}$/HCN and HNC/HCN maps have values less than one, consistent with the lower resolution, single-dish results of \citet{b219}. We propose that NGC~1808 is PDR heated. The ratio values of NGC~1808 are consistent with the intermediate PDR class of \citet{loenen2008}. \\

The detection of three PDRs is significant as very few XDRs and PDRs have been identified. These results demonstrate that in NGC~1097, NGC~1365 and NGC~1808 which host PDRs, the vigourous star formation has the dominant effect on the circum-nuclear molecular gas. Our PDR classification using the integrated intensity ratios of HCN, HCO$^{+}$ and HNC reproduces the literature classification of NGC~1365 arrived at through alternative methods, demonstrating the validity and usefulness of these molecular lines in the identification of these regions. Interestingly we find no galaxy which shows the PDR signature in only one part of the observed nuclear region. This lack of variation can only be deduced from interferometric data and is a major advantage of using the ATCA for this study as opposed to a single dish telescope. \\	

\subsection{Rotation curves and dynamical masses}
\label{sec:Dynamical Masses}
Rotation curves have been derived from position-velocity diagrams (PVDs) for our galaxy sample and these are presented in Online Appendix D. The PVD and rotation curve of NGC 4945 HCN are presented in~\autoref{fig:example_plots} as an example. The PVDs were extracted from the data cubes of each molecular line along the kinematical major axis of each galaxy using the program kpvslice$^{8}$\let\thefootnote\relax\footnote{$^{8}$\href{http://www.atnf.csiro.au/computing/software/karma/user-manual/kpvslice.html}{http://www.atnf.csiro.au/computing/software/karma/user-manual/kpvslice.html}}. The position angles of the position-velocity (\emph{p-v}) cuts are listed in~\autoref{tab:Observed Enclosed Mass Estimates} alongside the dynamical mass estimates, and are overlaid on the moment zero maps (see Online Appendix A). Velocities and radii were extracted from the PVDs using an \textsc{IDL} code to produce the rotation curve. \\

There were no clear indications of rotation in the moment one maps (intensity weighted velocity maps, see Online Appendix A) of NGC~3256 or NGC~5236, therefore we exclude these sources from the dynamical mass calculation. Clear, well defined PVDs, suitable for the construction of rotation curves, could be produced for NGC~1097 (HCN), NGC~1365 (HCN, HCO$^{+}$, HNC), NGC~1808 (HCN, HCO$^{+}$, HNC) and NGC~4945 (HCN, HCO$^{+}$, HNC). \\

The dynamical mass represents the total mass of all matter interior to a particular radius, including contributions from the molecular gas, dust, stars and central black hole. It provides a useful characterisation of the circum-nuclear region of galaxies, giving an upper limit on the mass of the central black hole \mbox{\citep[e.g.][]{garcia1988}}. Our high resolution data means we can estimate the dynamical mass at small radii. \\

The relation between dynamical mass and the kinematics of the gas, assuming Keplerian rotation, is given for example by \mbox{\citet{b75}}:
\begin{equation}
\rm{M_{dyn}=232 \eta R {v_{R}}^{2}}
\label{eq:black_hole_mass}
\end{equation}

\noindent where \emph{M$_{dyn}$} is the dynamical mass in M$_{\astrosun}$, \emph{v$_{R}$} is the inclination (\emph{i}) corrected radial velocity in km\,s$^{-1}$ (~$v_{R}=v_{uncorrected}/sin(\emph{i})$~) at distance \emph{R} in pc from the centre of mass and $\eta$ is a constant with a value between zero, for the most flattened disk mass distribution, and one, for a spherical mass distribution. The moment one maps indicate that the circum-nuclear gas of NGC~1097 (HCN), NGC~1808 (HCN, HCO$^{+}$, HNC), and NGC~4945 (HCN, HCO$^{+}$, HNC) is distributed in a disk. We therefore assume an intermediate value of $\eta$=0.8, representing a disk-like mass distribution in between that of the most flattened disk and a sphere in order to perform the dynamical mass calculation. The gas of NGC~1365 (HCN, HCO$^{+}$, HNC) appears to trace a warped disk. The straight \emph{p-v} cut does not capture its full rotation, therefore NGC~1365 is excluded from the dynamical mass analysis.     \\

\begin{table*} 
\centering
\caption{\textbf{Dynamical mass estimates.} \emph{Column 1} lists the source name,  \emph{Column 2} provides the molecular species (J=1$\rightarrow$0 transition), \emph{Column 3} presents the kinematical position angle and \emph{Column 4} denotes the inclination angle used to convert the observed to real rotational velocities (the references for which are listed in~\hyperref[fig:table1]{Table 1}). \emph{Column 5} provides the radius, \emph{Column 6} lists the matching inclination corrected velocity at that radius, \emph{Column 7} presents the corresponding dynamical mass calculated using~\autoref{eq:black_hole_mass}. The radii were initially calculated in seconds of arc, and have been converted to parsec using the distances listed in~\hyperref[fig:table1]{Table 1}. For comparison \emph{Column 8} and \emph{Column 9} present the dynamical mass estimate (calculated at a comparable radius to that listed in \emph{Column 5}), and the corresponding radius available in the literature. These estimates were not made with the same molecular lines as this work. References: $^{a}$\citet{b187},
$^{b}$\citet{b244}.\\} 
\label{tab:Observed Enclosed Mass Estimates}

\begin{tabular}{@{}l l l l c c l l l@{}}
\toprule

\textbf{Source} & \multicolumn{1}{c}{\textbf{Mol.}} & \multicolumn{1}{c}{\textbf{PA}} & \multicolumn{1}{c}{\textbf{\emph{i}}} & \multicolumn{1}{c}{\textbf{R}} & \multicolumn{1}{c}{\textbf{V$_{\rm{R}}$}} & \multicolumn{1}{c}{\textbf{M$_{\rm{dyn}}$}}    & \multicolumn{1}{c}{\textbf{Lit. M$_{\rm{dyn}}$}} & \multicolumn{1}{c}{\textbf{Lit.} \textbf{R}} \\
 & & \multicolumn{1}{c}{\textbf{[$^{\circ}$]}}& \multicolumn{1}{c}{\textbf{[$^{\circ}$]}} & \multicolumn{1}{c}{\textbf{[pc]}} & \multicolumn{1}{c}{\textbf{[km\,s$^{-1}$]}} & \multicolumn{1}{c}{\textbf{[M$_{\astrosun}$]}} & \multicolumn{1}{c}{\textbf{[M$_{\astrosun}$]}} & \multicolumn{1}{c}{\textbf{[pc]}}  \\
\midrule
\noalign{\vskip 1mm}
NGC~1097	&	HCN	&	$\sim$141	& 34 &	70	$\pm$	40	&	360	$\pm$	5	&	(1.7\,$\pm$\,1.0)$\times$10$^{9}$		&			2.8$\times$10$^{8}$ $^{a}$	&	40$^{a}$		\\
NGC~1808	&	HCN	&	$\sim$139	& 50 &	53\,$\pm$\,30	&	80\,$\pm$\,4	&	(6.0\,$\pm$\,3.4)$\times$10$^{7}$		&	$-$		&		$-$		\\
	&	HCO$^{+}$	&	$\sim$138	&	& 160\,$\pm$\,30	&	115\,$\pm$\,4	&	(4.0\,$\pm$\,0.8)$\times$10$^{8}$		&				&			\\
	&	HNC	&	$\sim$139	& &	110\,$\pm$\,30	&	105\,$\pm$\,4	&	(2.3\,$\pm$\,0.6)$\times$10$^{8}$		&				&			\\
NGC~4945	&	HCN	&	$\sim$43	&	78 & 18\,$\pm$\,9	&	130\,$\pm$\,3	&	(5.9\,$\pm$\,0.3)$\times$10$^{7}$		&	3.0$\times$10$^{7}$ $^{b}$	&	19$^{b}$		\\
	&	HCO$^{+}$	&	$\sim$43	&	& 18\,$\pm$\,9	&	140\,$\pm$\,3	&	(6.8\,$\pm$\,0.3)$\times$10$^{7}$		&				&			\\
	&	HNC	&	$\sim$45	&	& 36\,$\pm$\,9	&	120\,$\pm$\,3	&	(1.0\,$\pm$\,0.3)$\times$10$^{8}$		&				&			\\
\bottomrule
\end{tabular}
\end{table*}

For NGC~1808 no dynamical mass for a comparable radius is available in the literature. These results therefore provide an important first estimate of the dynamical mass of the circum-nuclear region, of $\sim$4.0$\times$10$^{8}$\,M$_{\astrosun}$ at $\sim$160\,pc.\\

\section{Summary and Conclusions}
\label{sec:conclusions}
In a sample of nine starbursts, the dense circum-nuclear gas traced by HCN, HCO$^{+}$ and HNC has been characterised, with seven detections of HCN (1$-$0) and HCO$^{+}$ (1$-$0), and four detections of HNC (1$-$0). With our sample of starburst galaxies we find that:
\begin{enumerate}
\item The HNC rivals the intensity of the HCN in NGC~5236, contrary to expectations. This may be due to hidden masses of cold gas, ion-neutral dominated chemistry or enhancement through mid-IR pumping. We have excluded the possibility that this is due to XDR chemistry.  \\

\item We detect PDRs within the circum-nuclear region of NGC~1097, NGC~1365 and NGC~1808. We find no galaxy which shows the PDR signature in only one part of the observed nuclear region. \\

\item The dynamical mass of the inner regions of three of the galaxies has been calculated. Our results represent the first such estimate for NGC~1808. 
\end{enumerate}

These results provide a solid foundation for the use of dense circum-nuclear molecular gas in the study of the connection between starbursts and AGN. Understanding this relationship is paramount to the study of galaxy formation and evolution. \\

\section*{Acknowledgements}
We would like to thank the referee for their careful reading and constructive comments, which significantly improved this paper. The authors would like to thank D. Espada for his work on the original ATCA observations and for useful discussions that contributed to this work. The Australia Telescope Compact Array is part of the Australia Telescope National Facility which is funded by the Commonwealth of Australia for operation as a National Facility managed by CSIRO. This paper includes archived data obtained through the Australia Telescope Online Archive (\href{http://atoa.atnf.csiro.au}{http://atoa.atnf.csiro.au}). This project was supported by the Brother Vincent Cotter Award for Physics (UNSW). LVM has been supported by Grant AYA2011-30491-C02-01 co-financed by MICINN and FEDER funds, and the Junta de Andaluc{\'i}a (Spain) grants P08-FQM-4205 and TIC-114. WAB acknowledges the support as a Visiting Professor of the Chinese Academy of Sciences (KJZD-EW-T01). The research leading to these results has received funding from the European Community's Seventh Framework Programme (/FP7/2007-2013/) under grant agreement No 229517. This project has made use of the NASA/IPAC Extragalactic Database (NED) which is operated by the Jet Propulsion Laboratory, California Institute of Technology, under contract with the National Aeronautics and Space Administration. This research has also made use of NASA's Astrophysics Data System Bibliographic Services. Graphs presented in this work were prepared using the \textsc{python} \textsc{matplotlib} graphics package \citep{matplotlib}. This research made use of \textsc{astropy}, a community-developed core \textsc{python} package for Astronomy (Astropy Collaboration, 2013). This research also made use of \textsc{aplpy}, an open-source plotting package for \textsc{python} hosted at \href{http://aplpy.github.com}{http://aplpy.github.com}.

\twocolumn



\section*{Supplementary material}

## A  Moment maps

**Figure A1: Moment zero and moment one maps.** Presented in the left column are the moment zero (velocity integrated specific intensity) maps and in the right column are the moment one (intensity weighted velocity) maps of the molecular data. The moment one maps are displayed for emission in the moment one maps $\geqslant 3\sigma$ for each molecular line. Source name and molecular species (J=1→0 transition) are listed on each map. For reference the beam is plotted in white in the bottom left corner of the moment zero map. The white line on the moment zero maps represents the direction of the $p$-$v$ cut used in the production of the position-velocity diagrams (PVDs) in Appendix C.

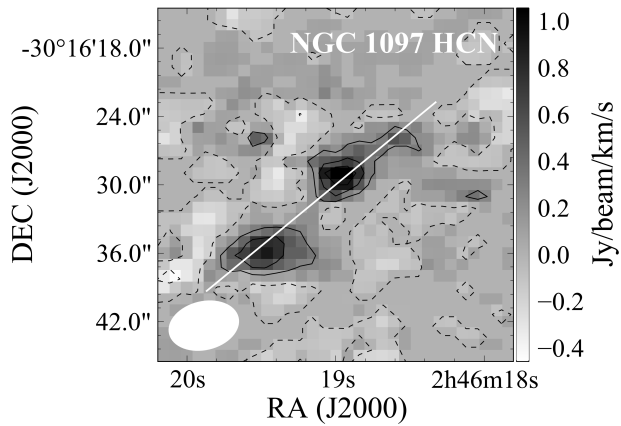

(i) **NGC 1097 HCN moment zero map.** The contours range from $-0.33$ to $0.92\,\mathrm{Jy\,beam^{-1}\,km\,s^{-1}}$ in increments of $0.31\,\mathrm{Jy\,beam^{-1}\,km\,s^{-1}}$.

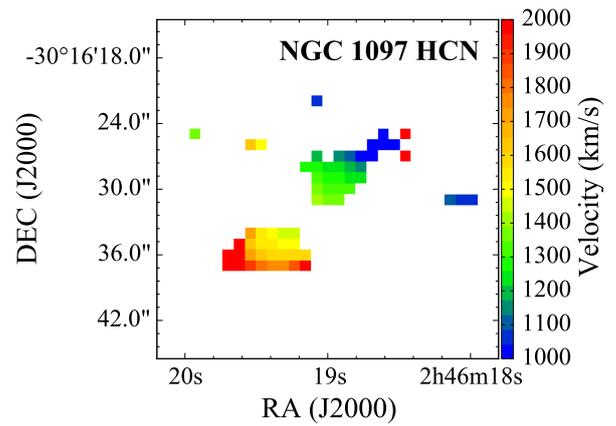

(ii) NGC 1097 HCN moment one map.

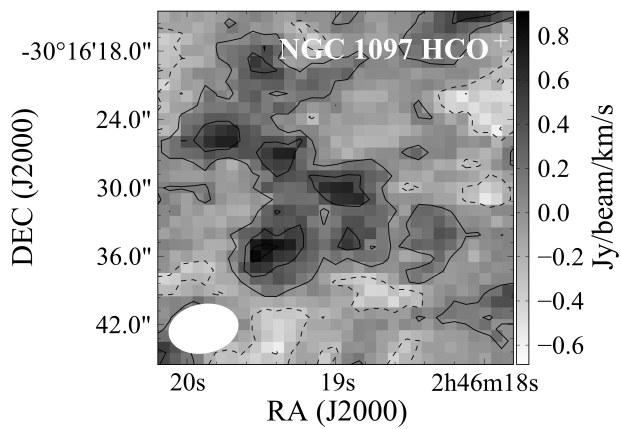

(iii) **NGC 1097 HCO$^+$ moment zero map.** The contours range from $-0.56$ to $0.77\,\mathrm{Jy\,beam^{-1}\,km\,s^{-1}}$ in increments of $0.33\,\mathrm{Jy\,beam^{-1}\,km\,s^{-1}}$.

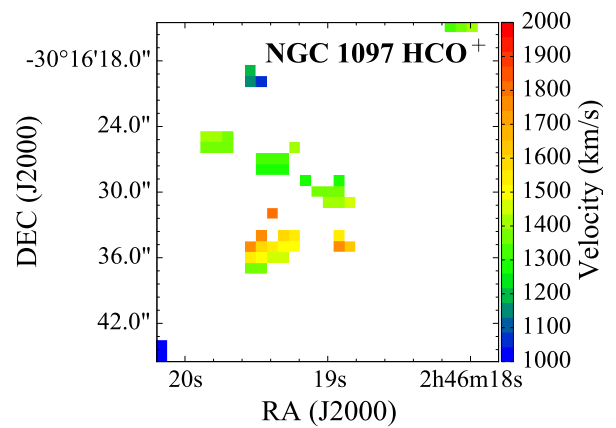

(iv) NGC 1097 HCO$^+$ moment one map.

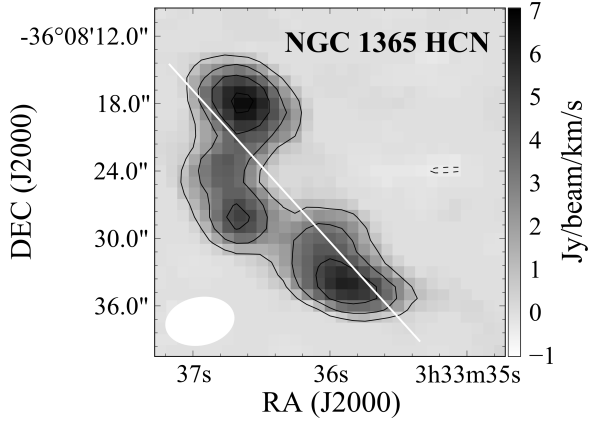

(v) **NGC 1365 HCN moment zero map.** The contours range from $-0.36$ to $6.32\,\mathrm{Jy\,beam^{-1}\,km\,s^{-1}}$ in increments of $1.67\,\mathrm{Jy\,beam^{-1}\,km\,s^{-1}}$.

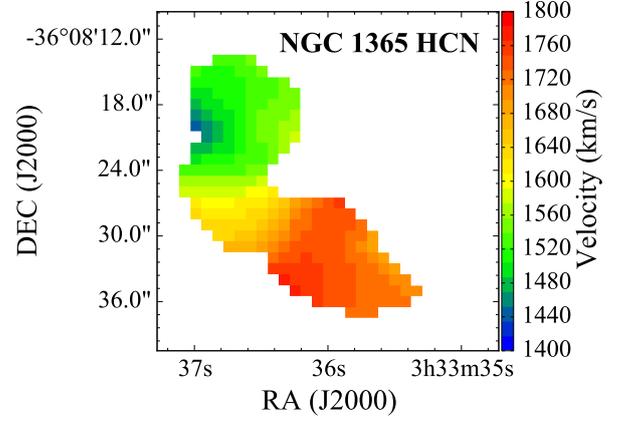

(vi) **NGC 1365 HCN moment one map.**

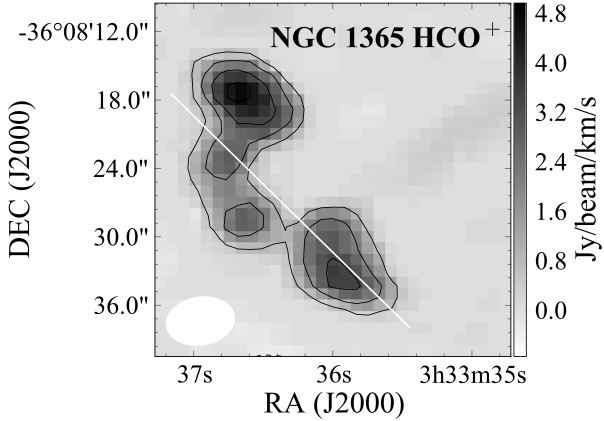

(vii) **NGC 1365 HCO$^+$ moment zero map.** The contours range from $-0.28$ to $4.44\,\mathrm{Jy\,beam^{-1}\,km\,s^{-1}}$ in increments of $1.18\,\mathrm{Jy\,beam^{-1}\,km\,s^{-1}}$.

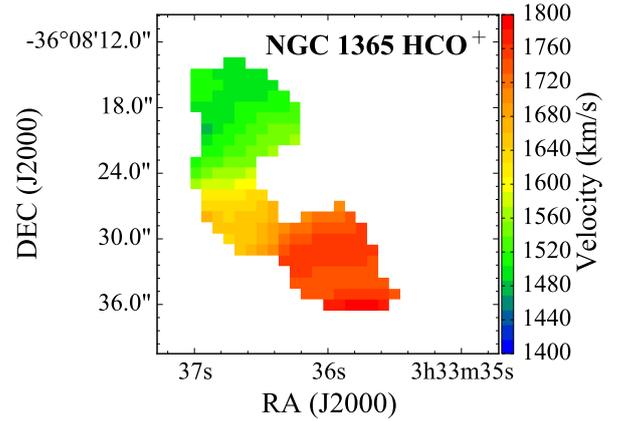

(viii) **NGC 1365 HCO$^+$ moment one map.**

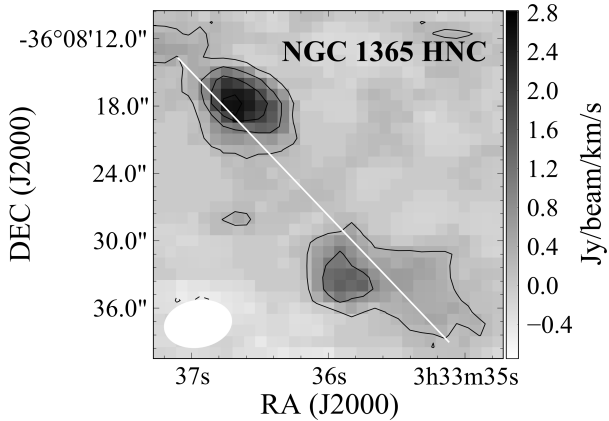

(ix) **NGC 1365 HNC moment zero map.** The contours range from $-0.45$ to $2.50\,\mathrm{Jy\,beam^{-1}\,km\,s^{-1}}$ in increments of $0.74\,\mathrm{Jy\,beam^{-1}\,km\,s^{-1}}$.

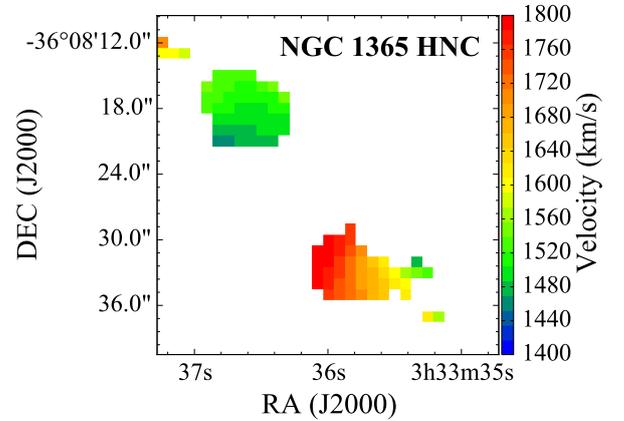

(x) **NGC 1365 HNC moment one map.**

**Figure A1:** *continued.*

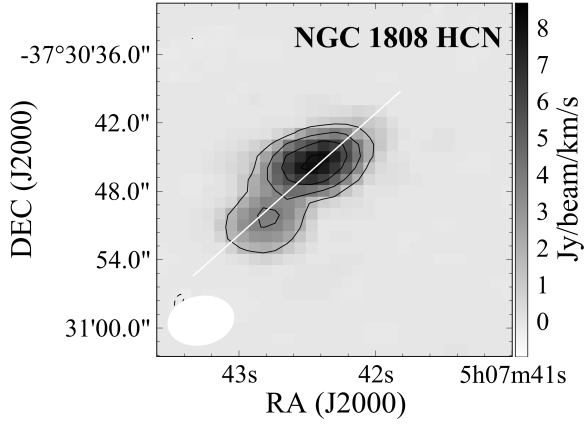

(xi) **NGC 1808 HCN moment zero map.** The contours range from $-0.16$ to $7.80\,\mathrm{Jy\,beam^{-1}\,km\,s^{-1}}$ in increments of $1.99\,\mathrm{Jy\,beam^{-1}\,km\,s^{-1}}$.

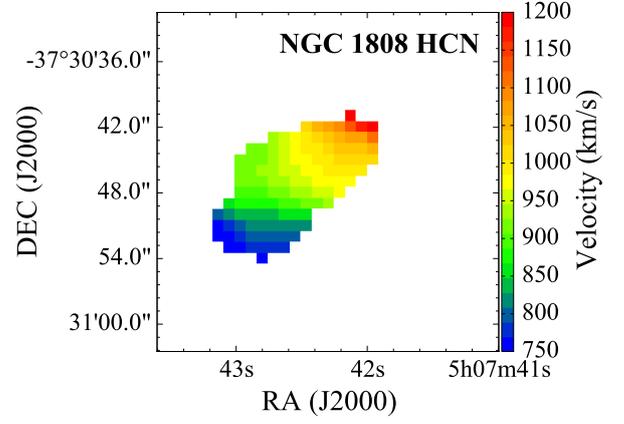

(xii) NGC 1808 HCN moment one map.

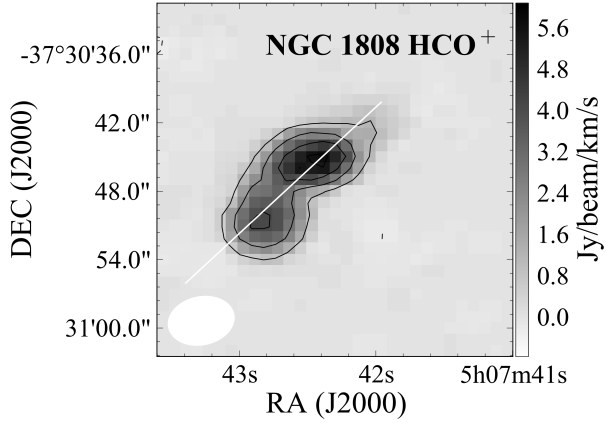

(xiii) **NGC 1808 HCO$^+$ moment zero map.** The contours range from $-0.20$ to $5.46\,\mathrm{Jy\,beam^{-1}\,km\,s^{-1}}$ in increments of $1.42\,\mathrm{Jy\,beam^{-1}\,km\,s^{-1}}$.

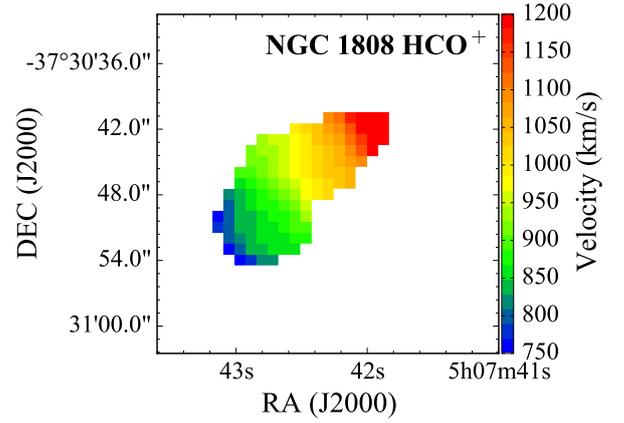

(xiv) NGC 1808 HCO$^+$ moment one map.

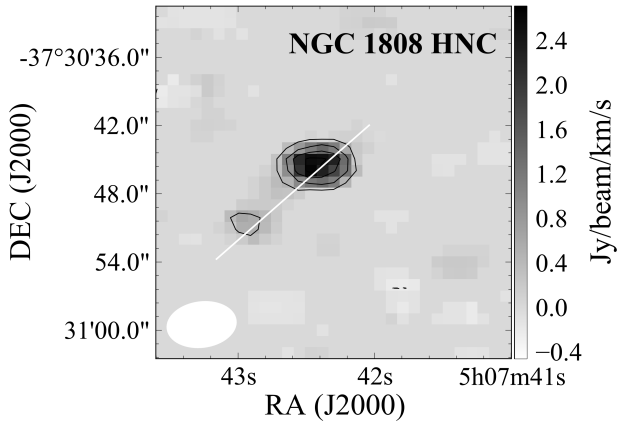

(xv) **NGC 1808 HNC moment zero map.** The contours range from $-0.20$ to $2.42\,\mathrm{Jy\,beam^{-1}\,km\,s^{-1}}$ in increments of $0.65\,\mathrm{Jy\,beam^{-1}\,km\,s^{-1}}$.

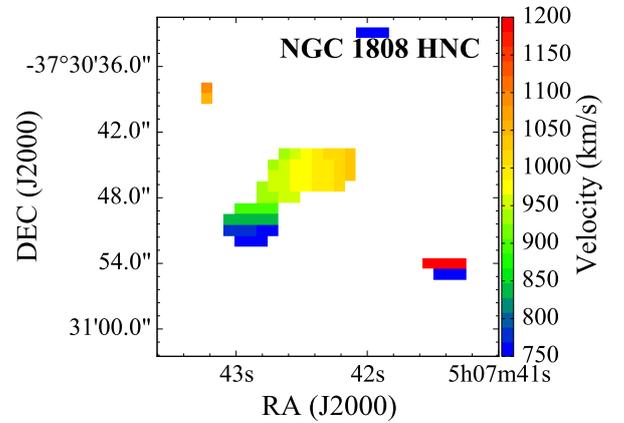

(xvi) NGC 1808 HNC moment one map.

**Figure A1:** *continued.*

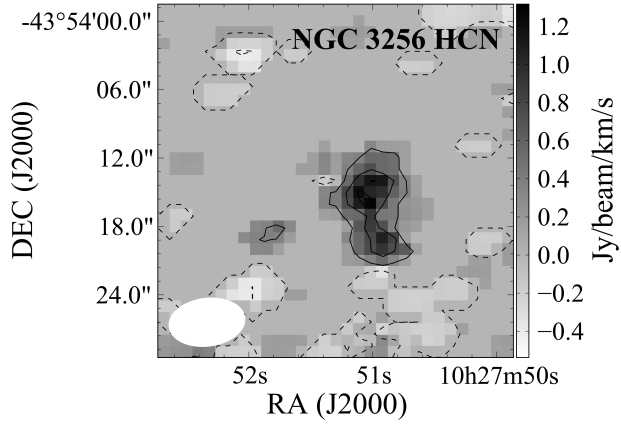

(xvii) **NGC 3256 HCN moment zero map.** The contours range from $-0.38$ to $1.15\,\mathrm{Jy\,beam^{-1}\,km\,s^{-1}}$ in increments of $0.38\,\mathrm{Jy\,beam^{-1}\,km\,s^{-1}}$.

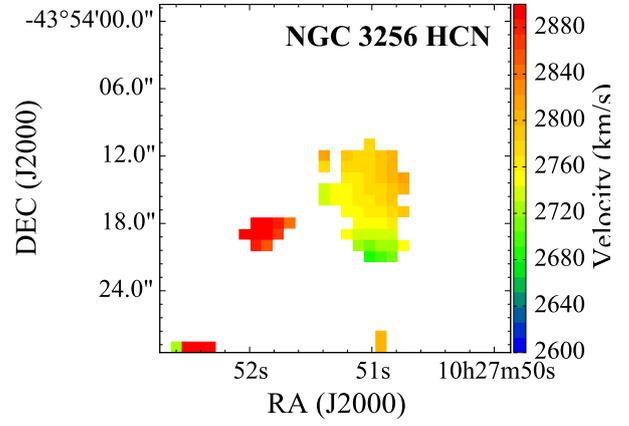

(xviii) **NGC 3256 HCN moment one map.**

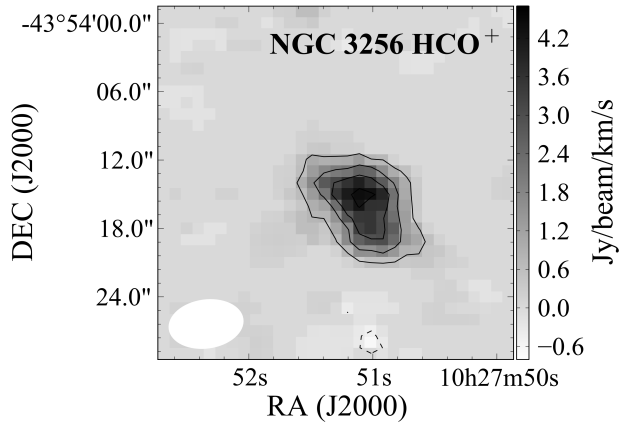

(xix) **NGC 3256 HCO$^+$ moment zero map.** The contours range from $-0.35$ to $4.2\,\mathrm{Jy\,beam^{-1}\,km\,s^{-1}}$ in increments of $1.14\,\mathrm{Jy\,beam^{-1}\,km\,s^{-1}}$.

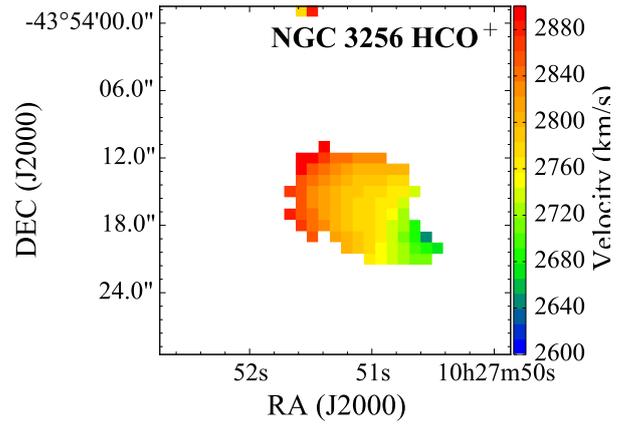

(xx) **NGC 3256 HCO$^+$ moment one map.**

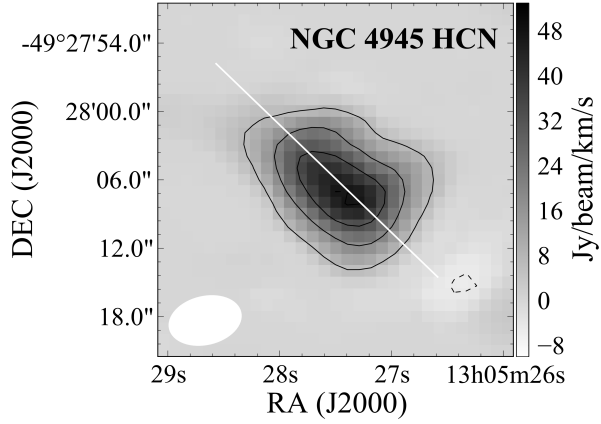

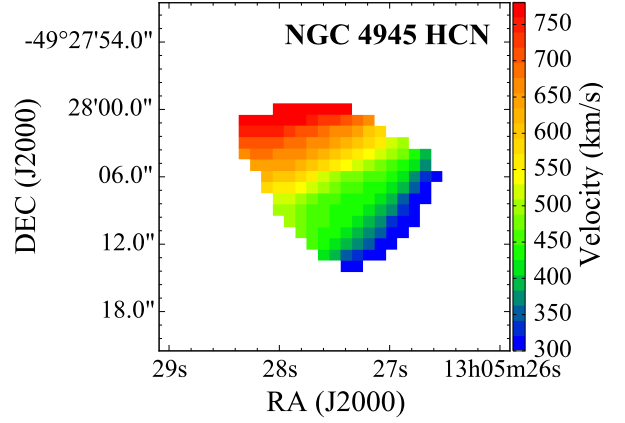

**(xxi)** NGC 4945 HCN moment zero map. The contours range from $-4.76$ to $47.24\,\mathrm{Jy\,Beam^{-1}\,km\,s^{-1}}$ in increments of $13\,\mathrm{Jy\,Beam^{-1}\,km\,s^{-1}}$.

**(xxii)** NGC 4945 HCN moment one map.

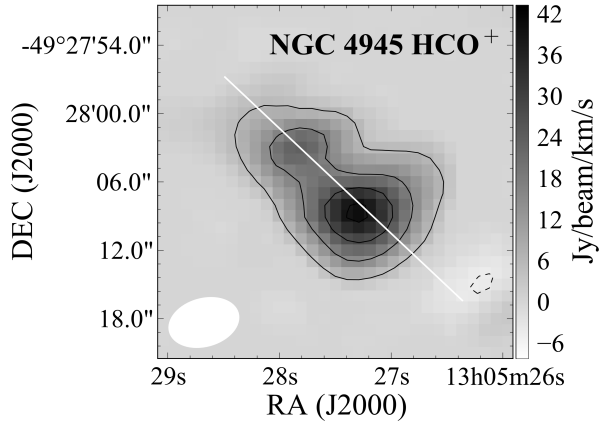

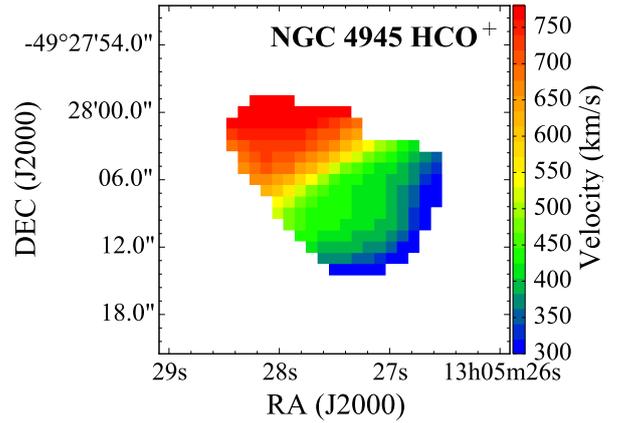

**(xxiii)** NGC 4945 HCO$^+$ moment zero map. The contours range from $-3.99$ to $38.56\,\mathrm{Jy\,Beam^{-1}\,km\,s^{-1}}$ in increments of $0.64\,\mathrm{Jy\,Beam^{-1}\,km\,s^{-1}}$.

**(xxiv)** NGC 4945 HCO$^+$ moment one map.

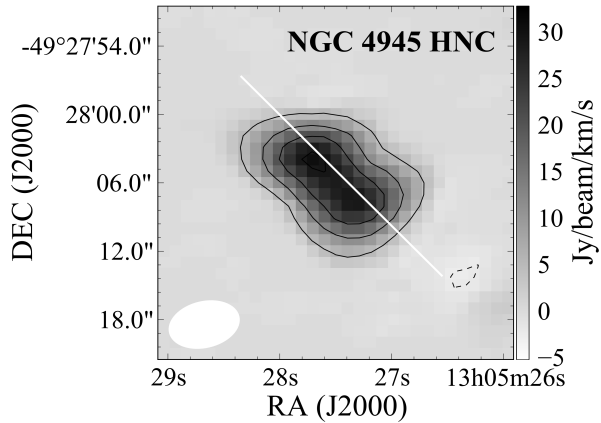

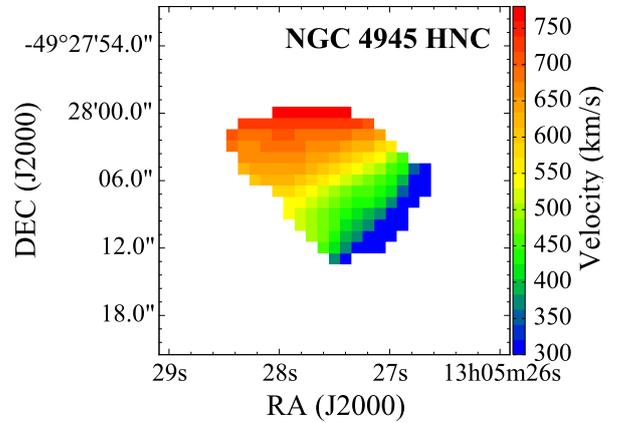

**(xxv)** NGC 4945 HNC moment zero map. The contours range from $-2.06$ to $29.34\,\mathrm{Jy\,Beam^{-1}\,km\,s^{-1}}$ in increments of $7.85\,\mathrm{Jy\,Beam^{-1}\,km\,s^{-1}}$.

**(xxvi)** NGC 4945 HNC moment one map.

**Figure A1:** *continued.*

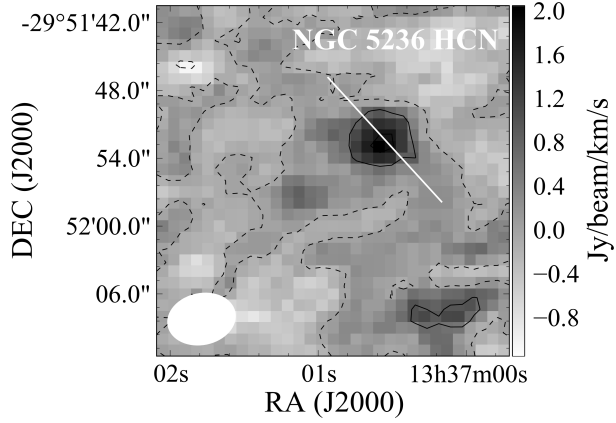

(xxvii) **NGC 5236 HCN moment zero map.** The contours range from $-0.89$ to $1.76\,\mathrm{Jy\,beam^{-1}\,km\,s^{-1}}$ in increments of $0.66\,\mathrm{Jy\,beam^{-1}\,km\,s^{-1}}$.

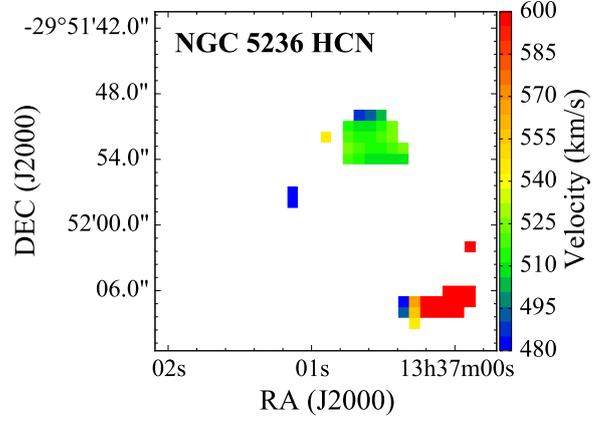

(xxviii) NGC 5236 HCN moment one map.

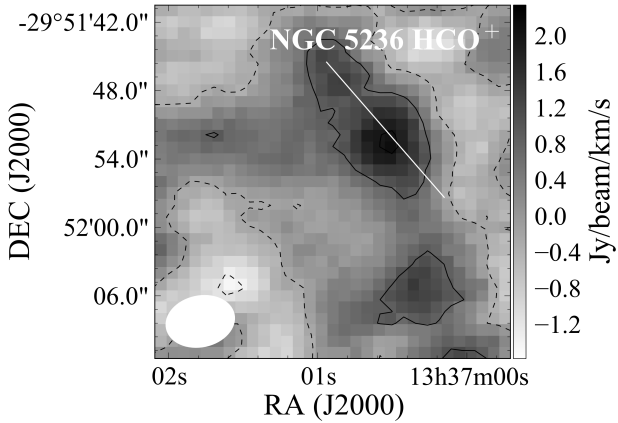

(xxix) **NGC 5236 HCO$^+$ moment zero map.** The contours range from $-1.25$ to $1.98\,\mathrm{Jy\,beam^{-1}\,km\,s^{-1}}$ in increments of $0.81\,\mathrm{Jy\,beam^{-1}\,km\,s^{-1}}$.

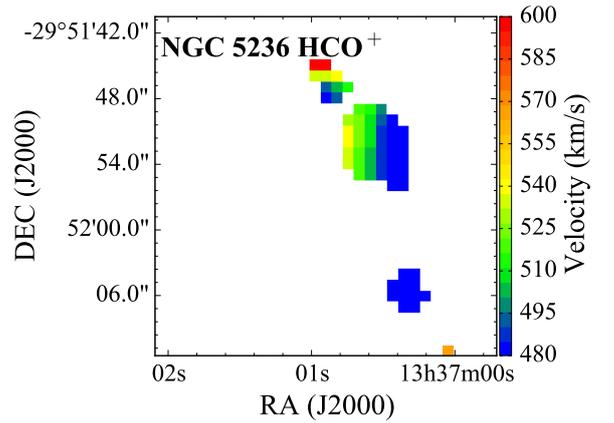

(xxx) NGC 5236 HCO$^+$ moment one map.

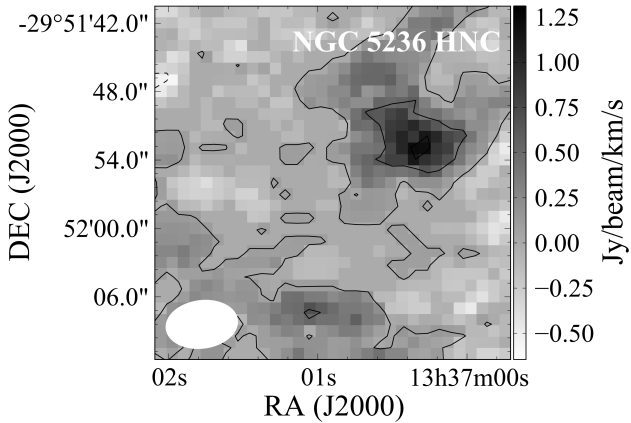

(xxxi) **NGC 5236 HNC moment zero map.** The contours range from $-0.48$ to $1.13\,\mathrm{Jy\,beam^{-1}\,km\,s^{-1}}$ in increments of $0.4\,\mathrm{Jy\,beam^{-1}\,km\,s^{-1}}$.

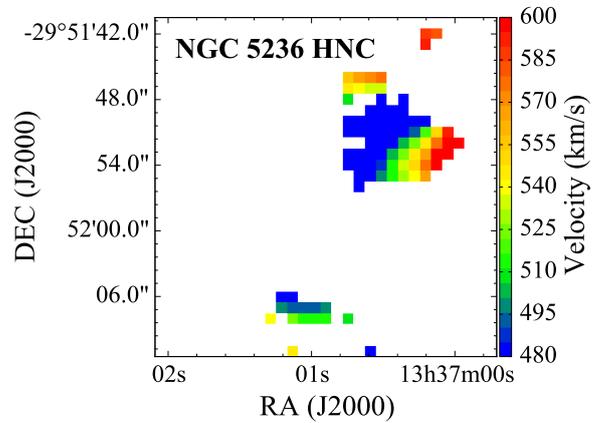

(xxxii) NGC 5236 HNC moment one map.

**Figure A1:** *continued.*

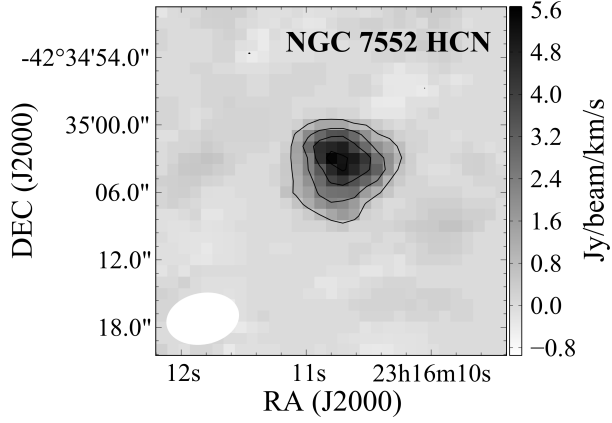

(xxxiii) **NGC 7552 HCN moment zero map.** The contours range from $-0.4$ to $5.06\,\mathrm{Jy\,beam^{-1}\,km\,s^{-1}}$ in increments of $1.37\,\mathrm{Jy\,beam^{-1}\,km\,s^{-1}}$.

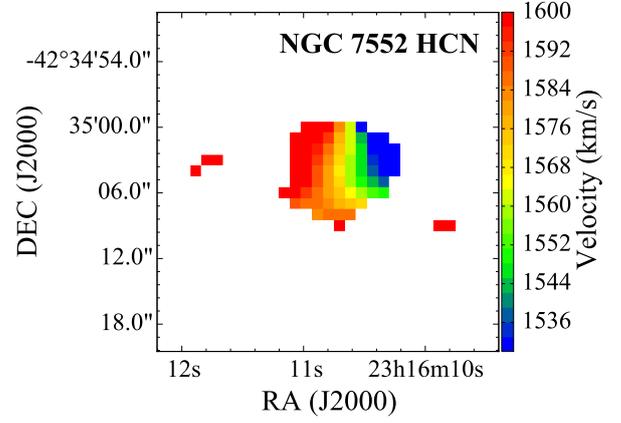

(xxxiv) **NGC 7552 HCN moment one map.**

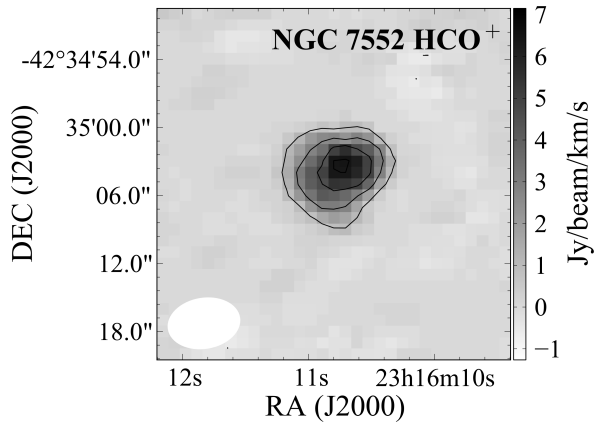

(xxxv) **NGC 7552 HCO$^+$ moment zero map.** The contours range from $-0.57$ to $6.39\,\mathrm{Jy\,beam^{-1}\,km\,s^{-1}}$ in increments of $1.74\,\mathrm{Jy\,beam^{-1}\,km\,s^{-1}}$.

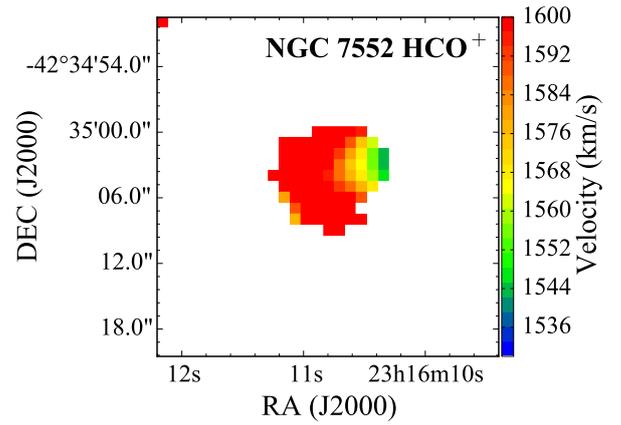

(xxxvi) **NGC 7552 HCO$^+$ moment one map.**

## B  'Missing flux' analysis

The data presented in this work was obtained with the ATCA interferometer. As with all interferometric measurements, there is some amount of emission on extended scales that is filtered out by the array. This was minimised by using the most compact array configuration available (H75), however, this 'missing flux' could potentially affect the integrated intensity ratios used in the classification of XDRs/PDRs. To determine the importance of this missing flux we estimate its magnitude here and calculate adjusted intensity ratios taking the missing flux into account. The details of this calculation are described, an example calculation presented, and the adjusted ratios are listed below.

To estimate the magnitude of the missing flux we examined one particular velocity in the ATCA interferometric and the SEST single-dish (Baan et al., 2008) spectra for each molecular species of each source. This velocity was selected as being that of the peak of the ATCA emission. The ATCA flux density at that velocity was subtracted from the SEST flux density. The difference in the flux density was then scaled by a factor representing the area in arcsecs$^2$ of the $\geqslant 3\sigma$ emission in that velocity slice divided by the area of the SEST beam (HPBW=57 arcsecs) to give an estimate of the missing flux density that would be 'hidden beneath' our ATCA detection. This process is summarised in the equation:

$$\text{Missing flux density} = \frac{\text{area of} \geqslant 3\sigma \text{ ATCA emission region (arcsec}^2)}{\text{area of SEST beam (arcsec}^2)} \times (\text{SEST} - \text{ATCA flux density (Jy), at a particlar velocity}) \tag{B1}$$

This missing flux density was then added to the ATCA flux density at that velocity and adjusted intensity ratios were calculated from these modified flux density values. For example for the HCO$^+$/HCN ratio (and similarly for the HNC/HCN ratio):

$$\text{Adjusted HCO}^+/\text{HCN ratio} = \frac{\text{ATCA HCO}^+ + \text{missing HCO}^+ \text{ flux density (Jy)}}{\text{ATCA HNC} + \text{missing HNC flux density (Jy)}} \tag{B2}$$

These adjusted ratios provide a first order approximation to assess the affect of the missing flux on the XDR/PDR classifications made in Section 3.2.2. They do not represent a better estimate to the integrated intensity ratios calculated in that section, but provide a check of their robustness. The data used for these calculations are listed in Table B1. The adjusted intensity ratio values are listed in Table B2, along with the original ATCA integrated intensity ratio values from Section 3.2.1. The difference between the original ATCA integrated intensity ratios and the adjusted ratios generally differ by less than 10% (with the exception of the NGC 3256 HCO$^+$/HCN ratio, see Table B2) and are consistent with the original ratios insofar as they remain either above or below one, which is the important consideration for XDR/PDR classifications. Such classifications made with the adjusted ratios are in agreement with the classifications made in Section 3.2.2. We therefore conclude that the missing flux does not present a significant problem in this work and that the XDR/PDR classifications made in Section 3.2.2 are valid.

Table B1: Missing flux calculation data. *Column 1* lists the source name, *Column 2* lists the molecular species (J=1−0 transition), *Column 3* provides the velocity at which the calculations were performed, *Column 4* gives the ATCA flux at that velocity, *Column 5* is the SEST flux (Baan et al. 2008) at that velocity, and *Column 6* represents the difference between the SEST and ATCA flux at that velocity. *Column 7* is the area of the ATCA emission $\geqslant 3\sigma$ at that velocity and *Column 8* provides an estimate of the missing flux calculated using the values in this table and Equation B1. All values are presented to 2 significant figures. Values were not rounded in the calculation of the ratios until the final step. Unfortunately no comparable SEST single dish data is available for NGC 1097.

| Source | Species | Velocity [kms$^{-1}$] | ATCA flux [Jy] | SEST flux [Jy] | Flux difference [Jy] | ATCA area [arcsec$^2$] | Missing flux [Jy] |
|---|---|---|---|---|---|---|---|
| **NGC 1365** | HCN | 1550 | 0.11 | 0.56 | 0.45 | 147 | 0.026 |
| | HCO$^+$ | 1550 | 0.09 | 0.38 | 0.29 | 146 | 0.016 |
| | HNC | 1550 | 0.05 | 0.34 | 0.29 | 73 | 0.0082 |
| **NGC 1808** | HCN | 950 | 0.086 | 0.34 | 0.25 | 63 | 0.0062 |
| | HCO$^+$ | 950 | 0.070 | 0.15 | 0.08 | 52 | 0.0016 |
| | HNC | 950 | 0.036 | 0.094 | 0.058 | 36 | 0.00082 |
| **NGC 3256** | HCN | 2750 | 0.075 | 0.13 | 0.056 | 88 | 0.0019 |
| | HCO$^+$ | 2750 | 0.13 | 0.23 | 0.12 | 67 | 0.0033 |
| **NGC 5236** | HCN | 500 | 0.032 | 0.56 | 0.53 | 42 | 0.0087 |
| | HCO$^+$ | 500 | 0.034 | 0.28 | 0.25 | 128 | 0.012 |
| **NGC 7552** | HCN | 1600 | 0.089 | 0.19 | 0.099 | 86 | 0.0033 |
| | HCO$^+$ | 1600 | 0.11 | 0.23 | 0.12 | 84 | 0.0039 |

Table B2: Adjusted intensity ratio results. *Column 1* is the source name. *Column 2* provides the adjusted intensity ratio values calculated with Equation B1 and Equation B2 and as described in text. *Column 3* lists the original ATCA integrated intensity ratio value as calculated in Section 3.2.1. Where two ratios are listed, the first refers to the ratio of the first (lowest velocity) Gaussian components of the relevant spectra, while the second listing refers to the ratio of the second (higher velocity) Gaussian components. *Column 4* presents the difference between the adjusted and original ratios as a percentage. Where two original ATCA ratios are listed, the percent difference is calculated from their average.

| Source | Adjusted ratio value | Original ATCA ratio value | Difference [%] |
|---|---|---|---|
| **HCO$^+$/HCN** | | | |
| NGC 1365 | 0.78 | 0.82 ± 0.06, 0.73 ± 0.06 | 0 |
| NGC 1808 | 0.78 | 0.86 ± 0.05 | 9 |
| NGC 3256 | 1.47 | 1.91 ± 0.22 | 23 |
| NGC 5236 | 1.13 | 1.08 ± 0.15 | 5 |
| NGC 7552 | 1.19 | 1.16 ± 0.15 | 3 |
| **HNC/HCN** | | | |
| NGC 1365 | 0.43 | 0.41 ± 0.04, 0.38 ± 0.04 | 4 |
| NGC 1808 | 0.40 | 0.38 ± 0.06 | 5 |

## C Intensity ratio maps

**Figure C1: Intensity ratio maps.** Presented in the left column are the HCO$^+$/HCN (1–0) intensity ratio maps and in the right column are the HNC/HCN (1–0) intensity ratio maps. The source name is listed on each plot. Plots are presented in pairs horizontally by source. The maps have been masked at the $\geqslant 3\sigma$ level of the HCN moment one map in Appendix A.

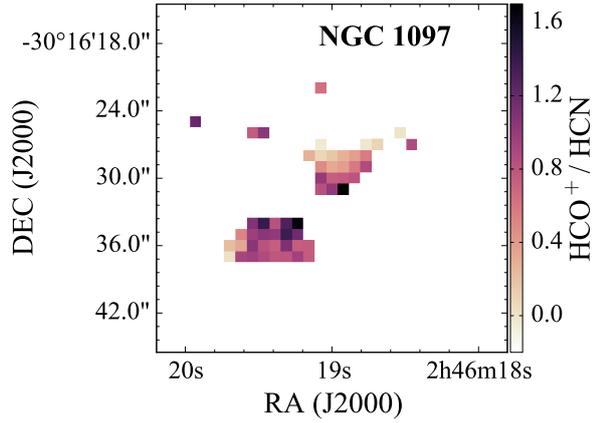

(i) NGC 1097 HCO$^+$/HCN intensity ratio map.

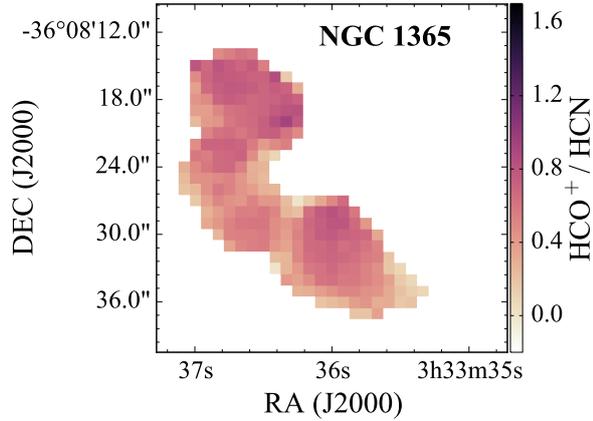

(ii) NGC 1365 HCO$^+$/HCN intensity ratio map.

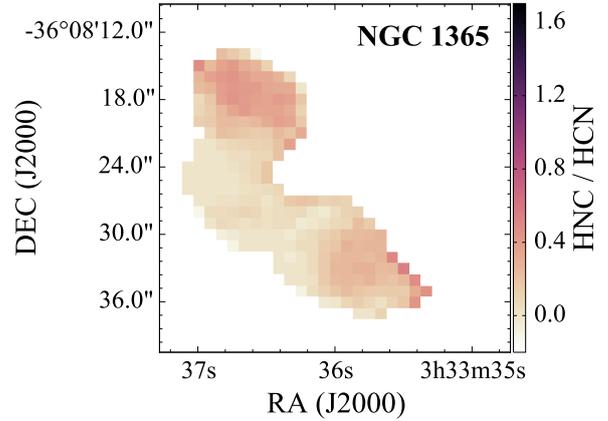

(iii) NGC 1365 HNC/HCN intensity ratio map.

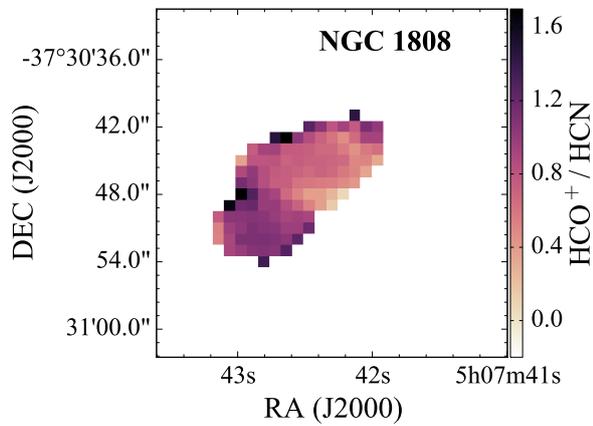

(iv) NGC 1808 HCO$^+$/HCN intensity ratio map.

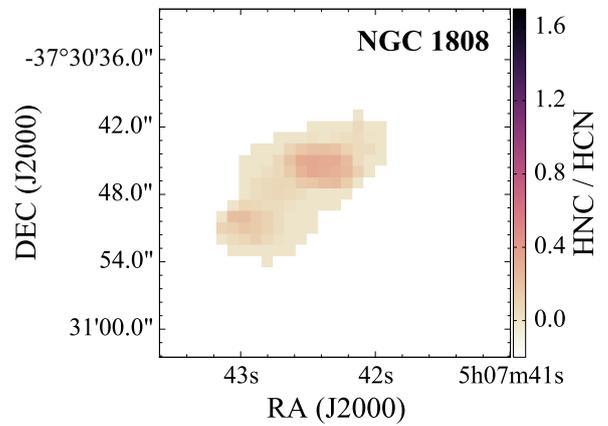

(v) NGC 1808 HNC/HCN intensity ratio map.

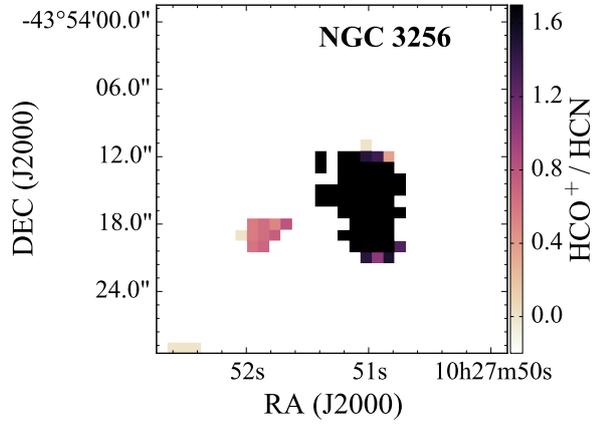

(vi) NGC 3256 HCO$^+$/HCN intensity ratio map.

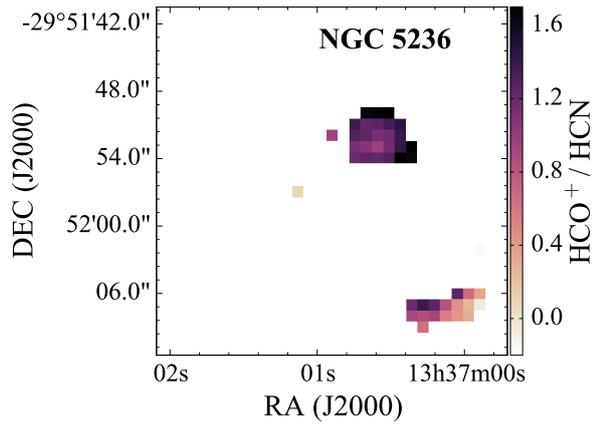

(vii) NGC 5236 HCO$^+$/HCN intensity ratio map.

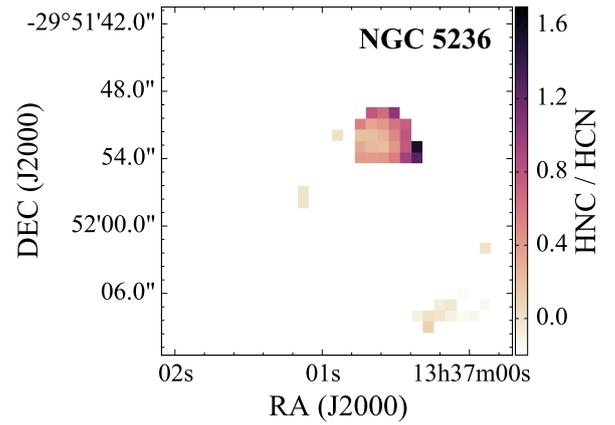

(viii) NGC 5236 HNC/HCN intensity ratio map.

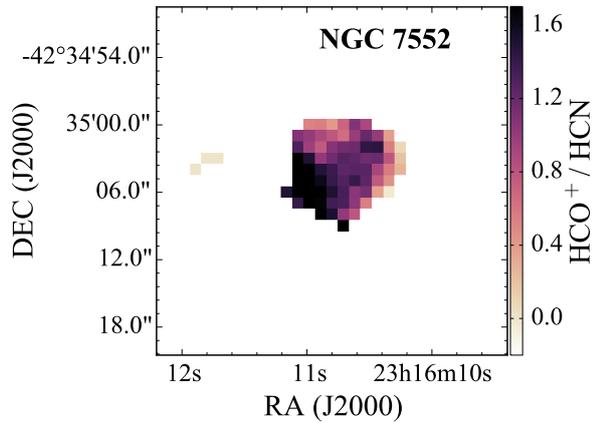

(ix) NGC 7552 HCO$^+$/HCN intensity ratio map.

## D  Position velocity diagrams and rotation curves

**Figure D1: PVDs and rotation curves.** Presented in the left column are the position-velocity diagrams (PVDs) and in the right column are the rotation curves of the molecular data. Source name and molecular line (J=1→0 transition) are listed on the plot. Plots are presented in pairs horizontally by source and molecular line. The $p$-$v$ cut used to produce the PVDs are presented as white lines on the corresponding moment zero (velocity integrated specific intensity maps) maps in Appendix A.

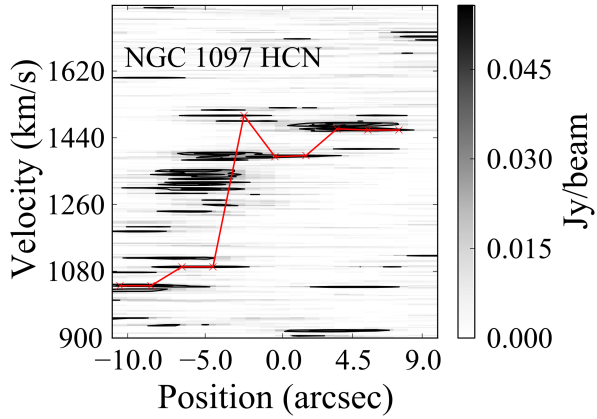

(i) NGC 1097 HCN position-velocity diagram and rotation curve.

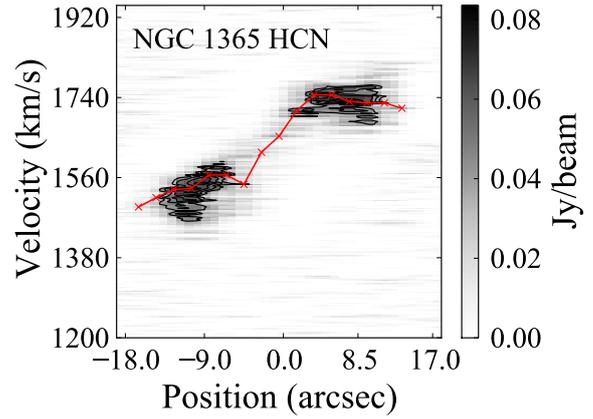

(ii) NGC 1365 HCN position-velocity diagram and rotation curve.

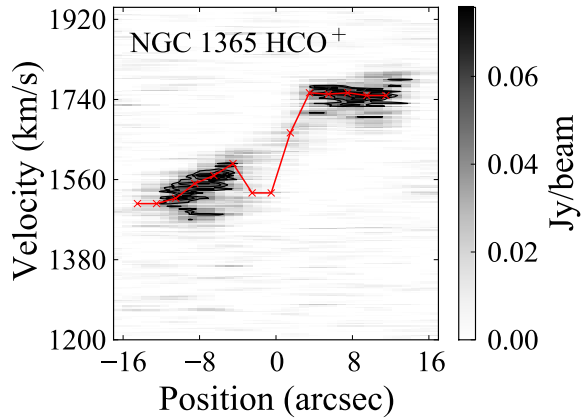

(iii) NGC 1365 HCO$^+$ position-velocity diagram and rotation curve.

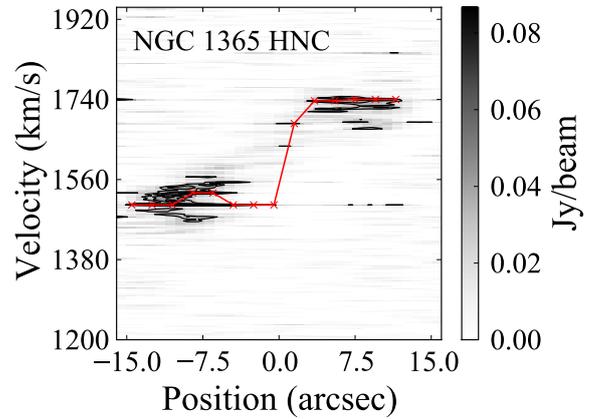

(iv) NGC 1365 HNC position-velocity diagram and rotation curve.

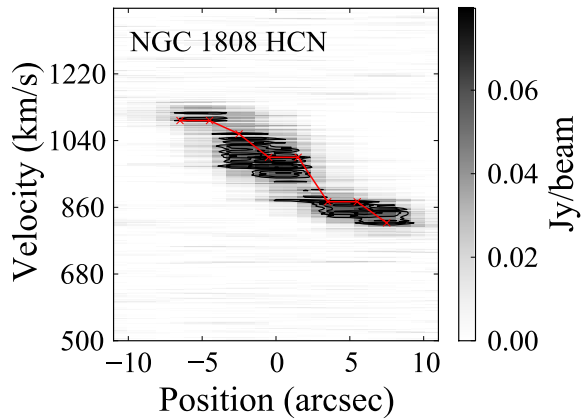

(v) NGC 1808 HCN position-velocity diagram and rotation curve.

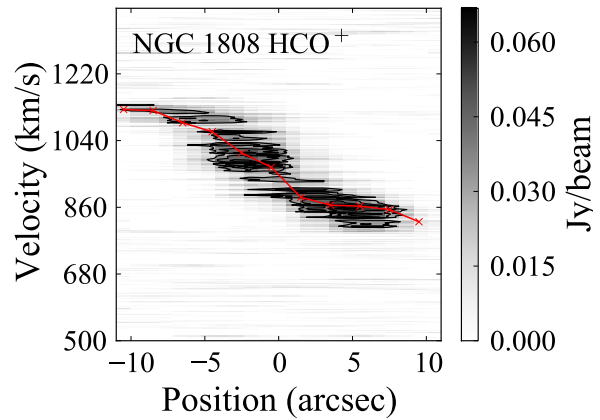

(vi) NGC 1808 HCO$^+$ position-velocity diagram and rotation curve.

**Figure D1:** *continued.*

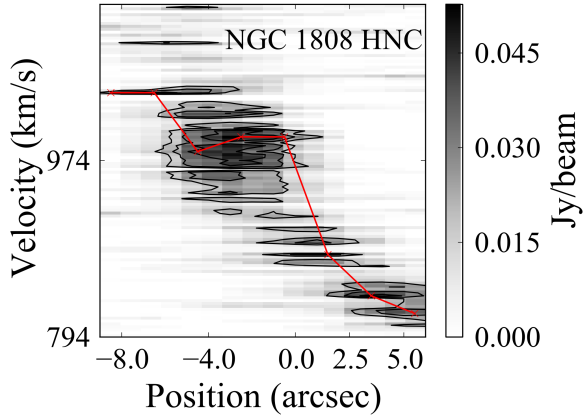

(vii) NGC 1808 HNC position-velocity diagram and rotation curve.

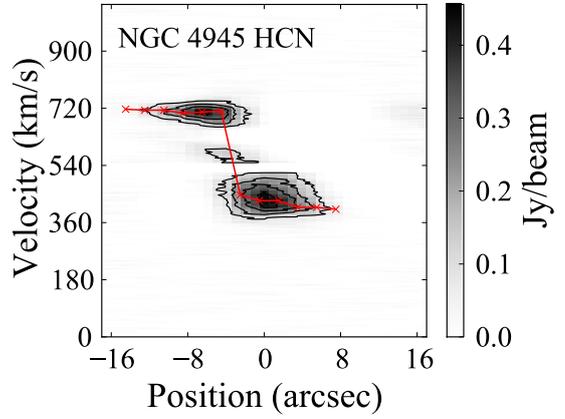

(viii) NGC 4945 HCN position-velocity diagram and rotation curve.

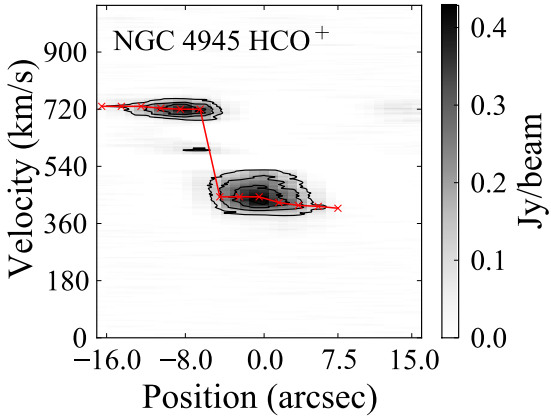

(ix) NGC 4945 HCO$^+$ position-velocity diagram and rotation curve.

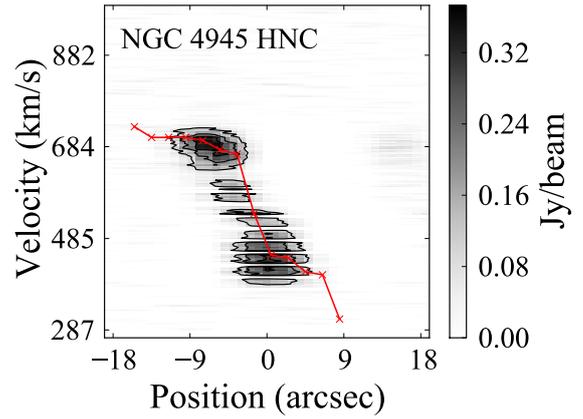

(x) NGC 4945 HNC position-velocity diagram and rotation curve.

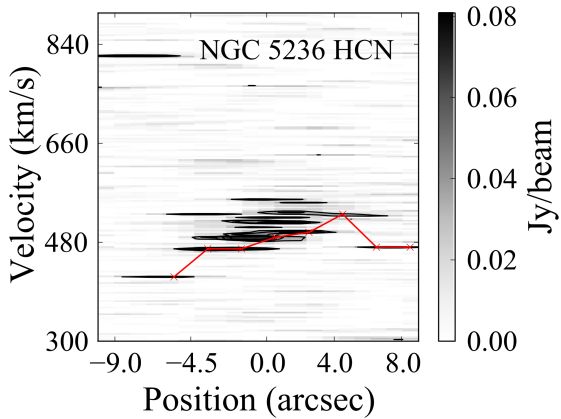

(xi) NGC 5236 HCN position-velocity diagram and rotation curve.

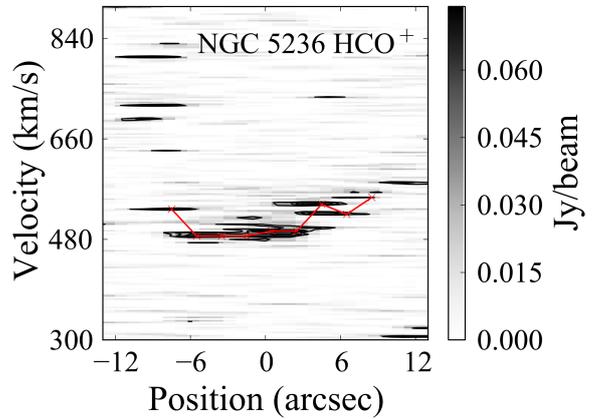

(xii) NGC 5236 HCO$^+$ position-velocity diagram and rotation curve.



\end{document}